\documentclass[aps,jcp,preprint,superscriptaddress]{revtex4}

\usepackage[utf8]{inputenc}

\usepackage{amsmath, amssymb}
\usepackage{tabularx,booktabs,array,dcolumn}
\usepackage{setspace}
\usepackage{graphicx,psfrag,subfigure}
\usepackage{xcolor}

\usepackage{marvosym}

\usepackage[all]{xy}

\usepackage{tikz}
\usetikzlibrary{calc}
\usetikzlibrary{patterns}

\usepackage{threeparttable}
\usepackage{rotating}
\usepackage{multirow}

\newcommand{\mx}[1]{\boldsymbol{#1}}
\newcommand{\bos}[1]{\boldsymbol{#1}}
\newcommand{\mr}[1]{\mathrm{#1}}

\newcommand{\pd}[2]{\frac{\partial #1}{\partial #2}}
\newcommand{\cm}{cm$^{-1}$}

\def\Eh{$E_\text{h}$}

\def\tr{^{\text{T}}}

\def\mel{m_\mr{e}}

\def\tlg{\tilde g}

\def\ra0{\mx{a}}

\def\som{Supplementary Material}

\def\htwo{\scalebox{0.5}{2}}

\def\mnuc{m_\text{n}}
\def\el{\text{el}}
\def\htwo{\hat{H}^\text{(2)}}
\def\ttwo{\hat{T}^\text{(2)}}

\def\tbo{\hat{T}^\text{(0)}}
\def\div{\text{div}}
\def\grad{\text{grad}}

\def\mmx{\mathcal{A}}
\def\massmx{\mathcal{M}}
\def\bmassmx{\bos{\mathcal{M}}}
\def\bxi{{\bos{\xi}}}

\def\tlg{\tilde{g}}

\def\div{\text{div}}
\def\grad{\text{grad}}

\def\nel{N_\text{el}}
\def\nnuc{N}

\def\mcr{\bos{\mathcal{R}}}
\def\naux{N_\text{aux}}

\def\Sgp{$\Sigma_\text{g}^+$}

\def\Pup{$\Pi_\text{u}^+$}

\def\jmxsph{\mx{J}_{\text{sp}}}
\def\omxsph{\mx{O}_{\text{sp}}}
\def\smxsph{\mx{S}_{\text{sp}}}

\def\jmxdiat{\mx{J}}
\def\omxdiat{\mx{O}}
\def\smxdiat{\mx{S}}

\def\massnuc{m_{\text{n}}}
\def\massel{m_{\text{el}}}
\def\tmassnuc{\tilde{m}_{\text{n}}}
\def\tmassvib{\tilde{m}_{\text{vib}}}
\def\tmassrot{\tilde{m}_{\text{rot}}}

\def\tXdSgp{\tilde{X}\ {^2\Sigma}_{\text{g}}^+}

\def\eel{E^{\text{(el)}}}
\def\enuc{E^{\text{(n)}}}
\def\hel{\hat{H}^{\text{(el)}}}

\def\papertwo{Paper~II}

\begin{document}

\title{%
Non-adiabatic mass correction 
to the rovibrational states of molecules.
Numerical application for the H$_2^+$ molecular ion
}

\author{Edit Mátyus}
\affiliation{Institute of Chemistry, Eötvös Loránd University, Pázmány Péter sétány 1/A, Budapest, H-1117, Hungary}

\date{\today}

\begin{abstract}
\noindent %
General transformation expressions of the second-order non-adiabatic Hamiltonian of the atomic nuclei,
including the kinetic-energy correction terms, 
are derived upon the change from laboratory-fixed Cartesian coordinates to 
general curvilinear coordinate systems commonly used in rovibrational computations.
The kinetic-energy or so-called ``mass-correction'' tensor elements are computed with
the stochastic variational method and floating explicitly correlated
Gaussian functions for the H$_2^+$ molecular ion in its ground electronic state. 
(Further numerical applications
for the $^4$He$_2^+$ molecular ion are presented in the forthcoming paper, \papertwo.) 
The general, curvilinear non-adiabatic kinetic energy operator expressions are used in the examples 
and non-adiabatic rovibrational energies and corrections are determined 
by solving the rovibrational Schrödinger equation including the diagonal Born--Oppenheimer 
as well as the mass-tensor corrections.

\end{abstract}

\maketitle

%
%
\section{Introduction}
\noindent%
The mass (or better: kinetic-energy) correction terms in the second-order non-adiabatic 
rovibrational Hamiltonian has been re-discovered in several very different contexts since
the advent of quantum mechanics.
In contrast to the diagonal Born--Oppenheimer (DBOC) correction, 
the mass-correction terms have been rarely included in
rovibrational computations, apart from the common arguments of using 
atomic instead of nuclear masses. The practice of using atomic masses is confirmed by 
\emph{a posteriori} by obtaining a better agreement with the experimentally observed
rovibrational transitions (with or
without accounting for any relativistic and radiative effects in the computations).

The earliest account of the kinetic-energy correction terms which we mention here is
from 1964 by Fisk and Kirtman  \cite{FiKi64} 
who wrote about a velocity-dependent effective potential energy 
surface (PES), which they obtained obtained using van Vleck's perturbation theory approach. 
Bunker and Moss derived effective non-adiabatic rovibrational Hamiltonians using van Vleck's perturbation
theory for di- \cite{BuMo77} and triatomic \cite{BuMo80} molecules and 
used their formalism for diatomics \cite{BuMcMo77}.
Later Schwenke elaborated on these correction terms and computed them in curvilinear coordinates 
for diatomic molecules \cite{Sch01H2p} as well as for the triatomic H$_2$O molecule \cite{Sch01H2O}. 

Pachucki and Komasa introduced non-adiabatic perturbation theory \cite{PaKo08} 
and arrived at the same mass-correction functions for diatomic molecules, 
which they successfully used for a series of systems \cite{PaKo09,PaKo12}, including
the H$_2$ molecule in its ground electronic state \cite{PaKo09}. 
Kutzelnigg \cite{Ku07} and Jaquet and Kutzelnigg \cite{JaKu08} derived and computed
mass-correction functions by starting out from a careful consideration of the
separation of the center of mass and the total mass of the molecule in an adiabatic theory.
There were also more empirical proposals for the ``rotational'' and ``vibrational''
mass correction functions of di- and triatomic molecules, which connected the mass correction to
the electron density assignable to the nuclei \cite{DiMoAl13}.

In an elegant and formal series of work, Teufel and co-workers introduced 
adiabatic perturbation theory in quantum dynamics and used it for the Born--Oppenheimer (BO) 
separation of the electronic and nuclear degrees of freedom, the coupling of which can be characterized
by the square root of the electron-to-nucleus mass ratio. The authors identified 
the almost invariant subspace for the electronic problem and using this subspace 
they derived an effective Hamiltonian for the quantum mechanical motion of the atomic nuclei
corresponding to an isolated electronic state (separated with gaps from all other electronic states). 
Expanding this effective Hamiltonian in terms of increasing orders of the coupling parameter, 
corrections are obtained to the (zeroth order) BO Hamiltonian of the atomic nuclei.
The second-order Hamiltonian
contains the DBOC and mass-correction tensor obtained also in other perturbative procedures.
Teufel and co-workers mention the derivation of Weigert and Littlejohn from 1993 \cite{WeLi93} 
using Weyl calculus for formally diagonalizing the multicomponent
wave equation, an approach which resulted a similar effective Hamiltonian for the nuclear motion. 

Most recently, independent of earlier work, the mass correction tensor \cite{prx17}
was derived from exact factorization through nuclear velocity perturbation theory \cite{ScAgSeGrVu15}, 
which successfully provides a theoretical framework also for vibrational circular dichroism. 

The relation of the mass-correction tensor to the computation of magnetic properties
has been observed (for diatomics) also by Bunker and Moss \cite{BuMo77}. 
Herman and Asgharian \cite{HeAs66} and Herman and Ogilvie \cite{HeOg98} 
pointed out its connection to the electronic contribution to 
the rotational and vibrational (paramagnetic) $g$-factors (see also Ref.~\cite{Og98book}).
For example, the $g$-factors computed using a linear response method with 
full-CI/aug-cc-pVTZ by Sauer et al. \cite{SaJeOg05}
for HeH$^+$ were found to be in a few \% agreement with 
computations of Pachucki and Komasa \cite{PaKo12} using
the optimized explicitly correlated Gaussian functions (ECGs).
Ogilvie and co-workers \cite{BaSaOdOg05} (as well as Pachuchki and Komasa \cite{PaKo09}) computed
the vibrational and rotational $g$-factors and adiabatic corrections for the hydrogen molecule.
The vibrational $g$-factor was computed for the bending mode of HCN in Ref.~\cite{BrReRu96}

Obviously, there has been a substantial theoretical and computational progress 
over the past half a century
concerning the theory and computational applications of 
the \emph{complete} second-order rovibrational Hamiltonian, 
which includes the mass-correction tensor. 
We find it fascinating that the same quantity, \emph{i.e.,} 
the same mass-correction term, appear in a variety of contexts and 
essentially the same quantity has been obtained starting from various
directions and by using very different (perturbational) formalisms. 
It would be interesting to explore the various aspects and theoretical 
connections between the different derivations.

After recognizing all these earlier, somewhat parallel developments, 
it is interesting to notice that the mass-correction tensor is not routinely included in 
the nowadays numerically ``exact'' rovibrational computations, in which 
however the PES (almost routinely) includes the diagonal Born--Oppenheimer correction, 
which is also a second-order term in the non-adiabatic Hamiltonian 
(see for example, Ref.~\cite{Ku07,PaSpTe07}). 
The accuracy of the
present-day high-resolution (precision) spectroscopy measurements (see for example
Ref.~\cite{SeJaMe16} discussed in \papertwo\ \cite{Ma18He2p})
implies that one has to go beyond the empirical non-adiabatic corrections in which
the nuclear masses are arbitrarily replaced with some effective (constant, usually the atomic) masses 
(which are thought to account for some of the mass-correction tensor effects). In particular,
there are at least two important ``families'' of \emph{small effects} to be accounted for in the computations:
``non-adiabatic corrections'' and ``relativistic and radiative corrections''---both represent 
challenges for a rigorous theoretical description. 

We wonder why the computation and the use of the mass-correction tensor did not 
become routine in rovibrational studies. 
We think and agree with the authors of Ref.~\cite{prx17} 
that for a widespread applicability
it would be important to compute the mass-correction tensor in \emph{Cartesian coordinates}
(Schwenke in Ref.~\cite{Sch01H2O} also mentions this direction as possible further development for 
his curvilinear derivation).
In particular, the widespread and general computation of the 
DBOC has become possible by its formulation and computation 
in simple Cartesian coordinates following Handy and co-workers \cite{HaLe96,IoAmHa96,HaYaSc86} 
(instead of using the system- and coordinate-dependent form). 
The laboratory-frame Cartesian coordinate expression of the DBOC 
was later confirmed in a stringent numerical test by Cencek and Kutzelnigg \cite{CeKu97} and 
was explained in formal terms by Kutzelnigg \cite{Ku97}.
The DBOC is a scalar quantity, while the mass-correction factor is a tensor. 
As it was also discussed by Schwenke \cite{Sch01H2O}, 
the mass-correction tensor is an inherently more complicated mathematical object.
Nevertheless, it can be computed using Cartesian coordinates with a selected 
frame for the nuclei from the electronic energies and wave-functions
as it was demonstrated in Ref.~\cite{prx17} at the equilibrium structures of the H$_2$, H$_2$O, and 
CH$_3$OH molecules.

In what follows, we derive the rovibrational Hamiltonian including 
the mass-correction term starting from laboratory-frame Cartesian coordinates to 
a general curvilinear coordinate system, in the spirit of 
the numerical-kinetic energy operator approach used in the GENIUSH protocol \cite{MaCzCs09}.
For the electronic structure computations, (due to the lack of any widely available
electronic structure method) we employ our in-house developed
computer program, QUANTEN (QUANTum mechanical treatment of Electrons and atomic Nuclei), 
which uses the variational method and explicitly correlated Gaussian functions
to solve the Schrödinger equation. If all charges belong to the quantum system we solve
a pre-Born--Oppenheimer problem \cite{MaRe12,Ma13,Ma18}. 
For the present work, we extended QUANTEN for the case of fixed external charges 
to solve the electronic Schrödinger equation. In the second part of the paper,
we explain how the necessary adiabatic and non-adiabatic quantities are computed.
%

Applications for a variety of poly-atomic and poly-electronic 
molecules will become possible when the Cartesian
mass-correction tensor can be computed with an efficient electronic structure package
over a broad range of molecular configurations. The implementation of the mass-correction tensor 
computed in Cartesian coordinates in a general curvilinear rovibrational program 
(\emph{e.g.,} the GENIUSH program)
should be straightforward based on the expressions derived in the first part of the article.
To demonstrate the applicability of the expressions and to show numerical examples
we make the calculations explicit and give numerical results for the homonuclear diatomic H$_2^+$
molecular ion and further applications follow for $^4$He$_2^+$ 
in \papertwo\ \cite{Ma18He2p}.

\clearpage

\section{The second-order non-adiabatic Hamiltonian}
\noindent%
In this section, we re-iterate the second-order non-adiabatic Hamiltonian 
derived by Panati, Spohn and Teufel \cite{PaSpTe07}
and adjust some of the notation to that used in the derivation from exact factorization \cite{prx17}.
Both derivations have been carried out in laboratory-frame Cartesian coordinates,
and these general, $N$-atomic expressions provide the most convenient starting point for our work.
The second-order non-adiabatic Hamiltonian of the atomic nuclei corresponding to a single electronic state 
(in Hartree atomic units $\hbar=\mel=a_0=1$) is:
\begin{align}
  \htwo
  =
  -\sum_{i=1}^N\sum_{a}
    \frac{1}{2\mnuc}
    \partial_{R_{ia}}^2
  +\sum_{i,j=1}^N\sum_{a,b}
    \frac{1}{2\mnuc}
    \partial_{R_{ia}}
    \left[%
      \frac{1}{\mnuc} \mmx_{ia,jb}
    \right]
    \partial_{R_{jb}}
  +V
  +
  \sum_{i=1}^N
  \sum_a
  \frac{1}{\mnuc} U_{ia}
  \label{eq:htwo}
\end{align}
where $\partial_{R_{ia}}=\partial/\partial_{R_{ia}}$
is short notation for the partial derivative with respect to the 
laboratory-frame Cartesian coordinates, $R_{ia}$ ($a=X,Y,Z$).
The first and thrid term is the BO kinetic energy operator and the potential energy surface (PES), respectively. 
The latter equals to the electronic energy, $\eel$,
the eigenvalue of the electronic Schrödinger equation:
\begin{align}
  \hel\psi &= \eel\psi 
  \label{eq:elSch}
\end{align}
with 
\begin{align}
  \hel
  &=
  -\frac{1}{2}\sum_{i=1}^{\nel} \Delta_{x_i}
  + 
  \sum_{i=1}^{\nel} \sum_{j>i}^{\nel}
    \frac{1}{|\bos{x}_i-\bos{x}_j|}
  -
  \sum_{i=1}^{\nel} \sum_{k=1}^{\nnuc}
    \frac{Z_k}{|\bos{x}_i-\bos{R}_k|}
  +
  \sum_{k=1}^{\nnuc}\sum_{l>k}^{\nnuc}
    \frac{Z_k Z_l}{|\bos{R}_k-\bos{R}_l|}
\end{align}
for the nuclear configuration $\boldsymbol{R}_k\ (k=1,2,\ldots,N)$ (
$Z_k\ (k=1,2,\ldots,N)$ label the electric charge of the nuclei).

The correction terms, multiplied by the second power of $\sqrt{1/\massnuc}$ are 
\begin{align}
  U_{ia}
  =
  \frac{1}{2}  
    \left\langle%
      \partial_{R_{ia}}\psi|
      (1-\hat{P})
      \partial_{R_{ia}}\psi
    \right\rangle_\el,
\end{align}
which gives rise to the well-known diagonal Born--Oppenheimer correction (DBOC) to the PES, 
$U = \sum_{i=1}^N \sum_a U_{ia}/\mnuc$, 
while the correction tensor to the kinetic energy is
\begin{align}
  \mmx_{ia,jb}(\mx{R})
  =
  2  
  \left\langle%
    \partial_{R_{ia}}\psi|
    (\hel-\eel)^{-1}(1-\hat{P})
    \partial_{R_{jb}}\psi
  \right\rangle_\el , 
  \label{eq:masscorrtensor}  
\end{align}
where 
$\hat{P}=|\psi\rangle\langle\psi|$ is a projector to the $\psi$ 
electronic eigenstate.
We usually consider the ground electronic state, but in principle $\eel$ and $\psi$ 
can correspond to any isolated, electronically excited state 
which is separated with a gap from rest of the electronic states \cite{PaSpTe07}.

Note that we have introduced a factor of 2 in the expression of $\mmx$ to synchronize the notation
with Ref.~\cite{prx17}.
Panati, Spohn, and Teufel derive the formalism for identical nuclear masses, $\mnuc$. 
The equations can be generalized to different nuclear masses by assuming that 
$R_{ia}$ are mass-scaled coordinates, \emph{i.e.,} $R_{ia}=\sqrt{m_i/\mnuc}R'_{ia}$.

For later use, we define the second-order non-adiabatic kinetic energy operator
as the sum of the terms containing the differential operators in $\htwo$ as
\begin{align}
  \ttwo
  &=
  -\sum_{i=1}^N\sum_{a}
    \frac{1}{2\mnuc}
    \partial_{R_{ia}}^2
  +\sum_{i,j=1}^N\sum_{a,b}
    \frac{1}{2\mnuc}
    \partial_{R_{ia}}
    \left[%
      \frac{1}{\mnuc} \mmx_{ia,jb}
    \right]
    \partial_{R_{jb}}
  \nonumber \\
  &=
  -
  \sum_{i,j=1}^N\sum_{a,b}
    \frac{1}{2\mnuc}      
    \partial_{R_{ia}}
    \left[%
      \delta_{ia,jb}
      -
      \frac{1}{\mnuc} \mmx_{ia,jb}
    \right]
    \partial_{R_{jb}}
  \nonumber \\    
  &=
  -
  \sum_{i,j=1}^N\sum_{a,b}  
    \frac{1}{2\mnuc}  
    \partial_{R_{ia}}    
    \mathcal{M}_{ia,jb}
    \partial_{R_{jb}}, 
\end{align}
where the elements of the effective mass tensor have been defined as
\begin{align}
  \mathcal{M}_{ia,jb}
  =
    \delta_{ia,jb}
    -
    \frac{1}{\mnuc} \mmx_{ia,jb},
  \label{eq:masstensor}
\end{align}
and we refer to $\mmx_{ia,jb}$ as (elements of) the mass-correction tensor.

For later convenience, a ``condensed-index''
labeling is introduced for the vector and tensor quantities as:
\begin{align}
  R_{ia}\rightarrow R_I 
  \quad\text{and} \quad
  \mathcal{A}_{ia,jb} \rightarrow \mathcal{A}_{IJ} 
  \quad\text{and} \quad
  \mathcal{M}_{ia,jb} \rightarrow \mathcal{M}_{IJ} 
  \label{eq:condlabel}
\end{align}
with
$(ia)\rightarrow{I=3(i-1)+a}$, $(jb)\rightarrow{J=3(j-1)+b}$, and
$a,b\in\lbrace 1(X),2(Y),3(Z)\rbrace$.
The expanded $(ia)$ and the condensed $(I)$ indices
will be used in an interchangeable manner.

\section{Coordinates and transformations to describe the quantum mechanical motion of the atomic nuclei\label{ch:coortrfo}}
\noindent
We may write the BO kinetic energy operator of the atomic nuclei in a compact form with
constant masses as
\begin{align}
  \tbo
  =
  -\frac{1}{2\mnuc}
  \div\ \grad ,
  \label{eq:gentbo}
\end{align}
while the second-order non-adiabatic kinetic energy operator in a similarly compact
form is
\begin{align}
  \ttwo
  =
  -\frac{1}{2\mnuc} 
  \div\ \bos{\mathcal{M}}\ \grad,
  \label{eq:genttwo}
\end{align}
where $\bos{\mathcal{M}}$ is the coordinate-dependent matrix defined
in Eq.~(\ref{eq:masstensor}). At this starting point, 
all quantities are in laboratory-frame Cartesian coordinates (LFCC).

Rovibrational computations can be efficiently performed (see for example Refs.~\cite{Te16,Ca17}),
if the laboratory-fixed Cartesian coordinates are replaced 
with a physically motivated (curvilinear) coordinate set, $\bos{\xi}$.
This physically motivated set includes a set of internal coordinates that are
well suited to describe the internal motions (vibrations),
three orientation angles which describe the orientation of the body-fixed
frame with respect to the laboratory frame (rotations), 
and three coordinates which describe the translation of the center of mass (translations).
In the BO framework, Eq.~(\ref{eq:gentbo}), the mass corresponding 
to the overall translation of the nuclei
is the sum of the nuclear masses. The effective mass corresponding to
the overall translation in the non-adiabatic Hamiltonian, Eq.~(\ref{eq:genttwo}),
will be discussed in Section~\ref{ch:applications}. 

\subsection{General curvilinear coordinates}
In a BO (or at least, constant-mass computations), one has to re-write
$\tbo$ according to the 
$R_1,R_2,\ldots,R_{3N}\Rightarrow \xi_1,\xi_2,\ldots,\xi_{3N}$ 
coordinate transformation:
\begin{align}
  \tbo
  =
  -\frac{1}{2\mnuc}
    \div\ \grad 
  \quad \Rightarrow \quad 
  \tbo_{\xi}
  =&
  -\frac{1}{2\mnuc}
    \div_{\bos{\xi}}\ \grad_{\bos{\xi}} 
\end{align}
where $\div_{\bos{\xi}}$ and $\grad_{\bos{\xi}}$
are the divergence and the gradient in the new coordinates (no subscript means 
plain laboratory-fixed Cartesian coordinates).
General expressions for the curvilinear form, $\tbo_{\xi}$ are routinely used
in variational rovibrational computations \cite{Lu00,LaNa02,MaCzCs09}.

Concerning the transformation of $\ttwo$ to curvilinear coordinates,
one has to consider the transformation
\begin{align}
  \ttwo
  =
  -\frac{1}{2\mnuc}
    \div\ \bos{\mathcal{M}}\ \grad 
  \quad \Rightarrow \quad 
  \ttwo_{\xi}
  =
  -\frac{1}{2\mnuc}
    \div_{\bos{\xi}}\ \bos{\mathcal{M}}^{(\xi)}\ \grad_{\bos{\xi}} 
\end{align}
together with the transformation, 
$\bos{\mathcal{M}}\rightarrow \bos{\mathcal{M}}^{(\xi)}$.

In all earlier rovibrational computations which included
mass-correction terms, tailor-made $\ttwo$ non-adiabatic kinetic energy operators
have been derived corresponding to specific $\xi$ choices (\emph{i.e.,} diatomic molecules, or
triatomic Radau coordinates \cite{Sch01H2O}),
and the resulting coordinate-dependent mass coefficients have been computed 
from electronic-structure theory.

Our aim in the present paper is to derive a general curvilinear expression for the 
non-adiabatic kinetic energy operator starting from the laboratory-frame expressions.
It will provide us with general formulae not only for the transformation
of the differential operators but also for the transformation of the mass tensor.
Having all these transformation expressions at hand the mass-correction tensor
computed in plain Cartesian coordinates by electronic structure theory
will be straightforwardly applicable in rovibrational computations. Hence,
we would arrive in some sence to a generalization of Handy's method of DBOC
for the mass-correction tensor.
We may say that we arrived at a general expression in curvilinear coordinates
if all operators are expressed in terms of the metric tensor and/or the Jacobi tensor,
which are fundamental mathematical objects for the 
$R_1,R_2,\ldots,R_{3N}\Rightarrow \xi_1,\xi_2,\ldots,\xi_{3N}$ 
coordinate change (and are routinely evaluated in the GENIUSH program over a grid
to construct the curvilinear kinetic-energy operator terms during the computation).

In curvilinear coordinates, $\xi_{n}$ ($n=1,2,\ldots,D$, here $D=3N$) 
with the covariant metric tensor 
$g_{\mu\nu}=\delta_{ij} \pd{R^i}{\xi^\mu}\pd{R^j}{\xi^\nu}$, 
the divergence of an $\mx{F}$ vector field (using Einstein's summation notation) is 
\begin{align}
  \text{div}_\bxi \mx{F}
  =
  \tlg^{-1/2}
  \partial_\mu \tlg^{1/2} F^\mu
\end{align}
where 
$\partial_\mu=\partial/\partial \xi^\mu$ and
$\tilde{g}=\text{det} g_{\mu\nu}$. 
The gradient of a function $\phi$ in the new coordinates is 
\begin{align}
  \text{grad}_\bxi \phi
  =
  \partial^\mu \phi
  =
  g^{\mu\nu} \partial_\nu \phi, 
\end{align}
which includes the $g^{\mu\nu}$ contravariant metric tensor which is the inverse 
of the covariant metric tensor, $g_{\alpha\beta}g^{\beta\gamma}=\delta_\alpha^\gamma$.

\vspace{1cm}
\paragraph{BO kinetic energy operator in curvilinear coordinates}
Using this notation, we can re-write the differential operator in the kinetic energy operator
with constant masses as
\begin{align}
  \text{div}\ \text{grad}\ \phi
  =
  \tlg^{-1/2} \partial_\mu \tlg^{1/2} g^{\mu\nu} \partial_\nu\ \phi
  \label{eq:tcurv1}
\end{align}
with the normalization condition
\begin{align}
  \int \phi^\ast\phi\ \tlg^{1/2}\ \text{d}\xi_1 \ldots \text{d}\xi_D = 1, 
  \label{eq:norm1}  
\end{align}
and thus the corresponding volume element is 
\begin{align}
  \text{d}V
  =
  \tlg^{1/2} \text{d}\xi_1 \ldots \text{d}\xi_D  .
\end{align}

Following Podolsky's work \cite{Po28} we introduce wave functions normalized according
to 
\begin{align}
  \int \chi^\ast\ \chi\ \text{d}\xi_1 \ldots \text{d}\xi_D,
  \label{eq:norm2}  
\end{align}
and thus 
we can re-write the Schrödinger equation and the kinetic energy operator 
into a more symmetric form by first inserting $\phi=\tlg^{-1/4}\ \chi$ 
and then multiplying with $\tlg^{1/4}$ from the left: 
\begin{align}
  2(E-V)\ \phi
  &= 
  \tlg^{-1/2}\ \partial_\mu\ \tlg^{1/2}\ g^{\mu\nu}\ \partial_\nu\ \phi \\
  2(E-V)\ \tlg^{-1/4}\ \chi 
  &= 
  \tlg^{-1/2}\ \partial_\mu\ \tlg^{1/2}\ g^{\mu\nu}\ \partial_\nu\ \tlg^{-1/4}\ \chi 
  \\
  2(E-V)\ \chi
  &= 
  \tlg^{-1/4}\ \partial_\mu\ \tlg^{1/2}\ g^{\mu\nu}\ \partial_\nu\ \tlg^{-1/4}\ \chi .
  \label{eq:schpod}
\end{align}
The last equation, Eq.~(\ref{eq:schpod}), has been referred to 
as the Podolsky form of the kinetic energy operator in curvilinear coordinates: 
\begin{align}
  \hat{T}^{\text{(BO)}}_{\text{Pod}}
  &=
  \frac{1}{2\mnuc}
  \tlg^{-1/4} \partial_\mu \tlg^{1/2} g^{\mu\nu} \partial_\nu g^{-1/4} ,
  \label{eq:kinpod0} 
\end{align}
which is the form of the kinetic energy of the atomic nuclei implemented
in the GENIUSH program \cite{MaCzCs09,FaMaCs11} and used
in several rovibrational computations, \emph{e.g.,} in Refs.~\cite{SaCsAlWaMa16,SaCsMa17}.

\vspace{1cm}
\paragraph{Second-order non-adiabatic kinetic energy operator in curvilinear coordinates}
In a similar spirit, we re-write the ``$\div\ \bmassmx\ \grad $'' operator to
curvilinear coordinates, $\xi_\mu$ $(\mu=1,2,\ldots,D)$, with the $g_{\mu\nu}$ metric 
and $J^i_\alpha$ Jacobian tensors as:
\begin{align}
  \div\ \bmassmx\ \grad\ \phi
  &=
  \tlg^{-1/2}\ \partial_\mu\ \tlg^{1/2}\ \massmx^{\mu}_\nu\ \partial^\nu\ \phi \nonumber\\
  &=
  \tlg^{-1/2}\ \partial_\mu\ \tlg^{1/2}\ 
  g^{\mu\alpha}\ \massmx_{\alpha\beta}\ g^{\beta\nu}\ 
  \partial_\nu\ \phi \nonumber \\
  &=
  \tlg^{-1/2}\ \partial_\mu\ \tlg^{1/2}\ g^{\mu\alpha}\ 
  J^i_\alpha\ \massmx_{ij}\ J^j_\beta\ 
  g^{\beta\nu}\ \partial_\nu\ \phi .
\end{align}
Note that the $\massmx_{\alpha\beta}$ element of the $\bmassmx$ tensor 
corresponds to the $\alpha$th, $\beta$th components in curvilinear coordinates,
which is related to the $i$th, $j$th Cartesian components by
\begin{align}
  \massmx_{\alpha\beta}
  &=
  \pd{R^i}{\xi^\alpha} 
  \massmx_{ij} 
  \pd{R^j}{\xi^\beta},  
\end{align}
and the $\mx{J}$ Jacobian tensor collects the derivatives of the Cartesian coordinates
with respect to the curvilinear coordinates: $J^i_{\alpha}=\frac{\partial R^i}{\partial \xi^\alpha}$.

For later convenience, we introduce the short notation
\begin{align}
  \div\ \bmassmx\ \grad\ \phi
  &=
  \tlg^{-1/2}\ \partial_\mu\ \tlg^{1/2}\   
  \tilde{G}^{\mu\nu}
  \partial_\nu\ \phi .
  \label{eq:divMgradGtilde}
\end{align}
with the effective $\tilde{G}^{\mu\nu}$ matrix
which includes not only the mass-weighted metric tensor but also
the second-order non-adiabatic corrections to the kinetic energy operator 
(compare with Eqs.~(\ref{eq:masscorrtensor}) and (\ref{eq:masstensor}) and note
that the condensed-index labeling defined in Eq.~(\ref{eq:condlabel}) is used in this section):
\begin{align}
  \tilde{G}^{\mu\nu}
  &=
  g^{\mu\alpha}\ 
  J^i_\alpha\ \massmx_{ij}\ J^j_\beta\ 
  g^{\beta\nu} .
\end{align}
It is important to recognize that elements of the mass-correction tensor are computed from 
the electronic wave function with the atomic nuclei positioned in 
a certain body-fixed (BF) frame \cite{Sch01H2O}.
We denote the mass matrix (mass-correction matrix, see Eq.~(\ref{eq:masstensor})) 
corresponding to this BF frame with 
$\bar{\massmx}_{ij}$ ($\bar{\mmx}_{ij}$). 
In order to obtain the mass (correction) tensor in the laboratory-fixed (LF) frame, 
we have to account for the rotation from
the BF frame to the LF frame. Let us represent this rotation
with an $\mx{O}$ matrix, and thus the relation between the LF mass matrix 
(mass-correction matrix), $\bos{\massmx}$ ($\bos{\mmx}$),
and the BF mass matrix, $\bar{\bos{\massmx}}$ ($\bar{\bos{\mmx}}$), is
\begin{align}
  \massmx_{ij}
  =
  \left(%
    \mx{O}
    \bar{\bos{\massmx}}
    \mx{O}\tr
  \right)_{ij}
  \label{eq:massmxbflf}
\end{align}
with
\begin{align}
  \mmx_{ij}
  =
  \left(%
    \mx{O}
    \bar{\bos{\mmx}}
    \mx{O}\tr
  \right)_{ij} 
  \label{eq:mmxbflf}
\end{align}
and
\begin{align}
  \tilde{G}^{\mu\nu}
  &=
  g^{\mu\alpha}\ 
  J^i_\alpha\ 
    \left(%
    \mx{O}
    \bar{\bos{\massmx}}
    \mx{O}\tr
  \right)_{ij}\ 
  J^j_\beta\ 
  g^{\beta\nu} \\
  &=
  g^{\mu\alpha}\  
    \left(%
    \mx{J}\tr
    \mx{O}
    \bar{\bos{\massmx}}
    \mx{O}\tr
    \mx{J}
  \right)_{\alpha\beta}\ 
  g^{\beta\nu} ,
  \label{eq:gbarfull}
\end{align}
which completes the expression for the effective $\tilde{G}^{\mu\nu}$ matrix 
including the $\bar{\bos{\mmx}}$ mass correction tensor 
computed in electronic structure theory
with a certain embedding (BF) of the atomic nuclei.
Besides the mass-correction matrix values, the effective $\tilde{G}$ tensor 
contains the metric tensor and the Jacobi matrix elements---mathematical objects
defined by a general coordinate transformation, and the orientation matrix 
which defines the BF frame used to compute the mass-correction matrix elements.

Thereby, the complete non-adiabatic kinetic energy operator expression
in general curvilinear coordinates is
\begin{align}
  \hat{T}^{(2)} \phi
  &=
  -\frac{1}{2\mnuc}\ 
  \div\ \boldsymbol{\massmx}\ \grad\ \phi \nonumber \\
  &=
  -\frac{1}{2\mnuc}\ 
  \tlg^{-1/2} 
  \partial_\mu
  \tlg^{1/2}
  \tilde{G}^{\mu\nu}
  \partial_\nu
  \phi
  \label{eq:divMgrad}
\end{align}
with the volume element
$
  \text{d}V
  =
  \tlg^{1/2}
  \text{d}\xi_1\ldots\text{d}\xi_{D}.
$.
The effective $\tilde{\mx{G}}$ matrix in general curvilinear coordinates 
is calculated from the $\bar{\mathcal{A}}_{ij}$ body-fixed mass-correction tensor elements
as:
\begin{align}
  \tilde{G}^{\mu\nu}
  &=
  g^{\mu\alpha}
  (%
    \mx{S}
    \bar{\massmx}
    \mx{S}\tr
  )_{\alpha\beta}  
  g^{\beta\nu} 
  \nonumber \\
  &=
  \frac{1}{\massnuc}
  g^{\mu\alpha}
  (%
    \mx{S}
    \mx{I}
    \mx{S}\tr
  )_{\alpha\beta}  
  g^{\beta\nu}  
  -
  \frac{1}{\massnuc^2}
  g^{\mu\alpha}
  (%
    \mx{S}    
    \bar{\mx{\mmx}}
    \mx{S}\tr    
  )_{\alpha\beta}  
  g^{\beta\nu} , 
  \label{eq:geff}
\end{align}
where
\begin{align}
  \mx{S}
  =
  \mx{J}\tr\mx{O} .
  \label{eq:smxdef}
\end{align}
$\bos{J}$, $\tlg^{1/2}$, $g^{\mu\nu}$ is 
the Jacobi tensor, the Jacobi determinant, and 
the contravariant metric tensor of the coordinate transformation, respectively.
Furthermore, 
$\mx{O}$ is the rotation matrix which transforms $\bar{\bos{\mathcal{A}}}$ 
obtained from electronic structure theory in a selected body-fixed frame
(which may be different from the body-fixed frame corresponding to the 
new $\bos{\xi}$ coordinates of the rovibrational Hamiltonian)
to the laboratory-fixed frame.

The general expression, Eq.~(\ref{eq:divMgrad}),
can be re-written to the Podolsky form, following
the same reasoning as for constant masses in Eqs.~(\ref{eq:tcurv1})--(\ref{eq:kinpod}), 
by absorbing the $\tlg^{1/2}$ factor
in the operator:
\begin{align}
  \hat{T}^{(2)}_\text{Pod} 
  &=
  -\frac{1}{2\mnuc}\ 
  \tlg^{-1/4} 
  \partial_\mu
  \tlg^{1/2}
  \tilde{G}^{\mu\nu}
  \partial_\nu
  \tlg^{-1/4}   ,
\end{align}
for which the wave functions are normalized with the volume element:
\begin{align}
 \text{d}V_\text{Pod}=\text{d}\xi_1\ldots \text{d}\xi_D .
\end{align}

\section{Computation of the second-order correction terms using an explicitly correlated Gaussian basis set \label{ch:elstruct}}
\noindent 
%
\noindent 
We solve the electronic Schrödinger equation, Eq.~(\ref{eq:elSch}),
and to compute the mass-correction correction tensor
with the QUANTEN program 
using floating explicitly correlated Gaussian functions (fECG)
and the stochastic variational method \cite{SuVaBook98} with regular refinements as
implemented in Refs.~\cite{MaRe12,Ma13}.
Since the electronic problem with fixed nuclei does not have the full O(3)
rotation-inversion symmetry, which the pre-Born--Oppenheimer problem has, 
the integral expressions are considerably simpler and were obtained from
the integrals derived for the generator function
(see for example the Supplementary Material of Ref.~\cite{MaRe12}):
\begin{align}
  g(\mx{r};\mx{A},\mx{s})
  =
  \exp\left[
    -\frac{1}{2} \mx{r}\tr (\mx{A}\otimes \mx{I}_3)\mx{r}
    +\mx{s}\tr \mx{r}
  \right],
\end{align}
which is related to a floating explicitly correlated Gaussian functions (fECG)
centered at $\mcr\in\mathbb{R}^{3\nel}$ as
\begin{align}
  f(\mx{r};\mx{A},\mcr)
  &=
  \exp\left[%
    -\frac{1}{2} (\mx{r}-\mcr)\tr (\mx{A}\otimes \mx{I}_3) (\mx{r}-\mcr)  
  \right] 
  \nonumber \\
  &=
  \exp\left[%
    -\frac{1}{2} \mcr\tr (\mx{A}\otimes \mx{I}_3) \mcr
  \right]
  \times   
  %
  \exp\left[%
    -\frac{1}{2} \mx{r}\tr (\mx{A}\otimes \mx{I}_3) \mx{r}
    + \mcr\tr (\mx{A}\otimes \mx{I}_3) \mx{r}
  \right] , 
\end{align}
The first term in the product is a constant with respect to
the integration for the electronic coordinates,
and thus can be accounted for by simple multiplication.

The non-linear parameters of the fECG basis functions ($\mcr$ centers and $\mx{A}$ exponents)
are generated in a stochastic variational optimization procedure 
and they are regularly refined using Powell's method \cite{Po04}
to minimize the electronic energy, $E$ (similarly to the pre-BO approach \cite{MaRe12,Ma13}).
We have used the full point-group symmetry for the energy minimization, and have
exploited the idempotency of the symmetry projector, so its explicit
effect had to be calculated only for the ket functions. We have also exploited
the convenient transformation properties of fECGs under the effect of symmetry operators,
which can be translated to the transformation of the parameterization of an fECG \cite{MaRe12,Ma18}.

In order to generate a potential energy curve for a diatomic molecule 
(and other correction quantities), 
we have carried out a full basis-set optimization (energy minimization) 
at only a few nuclear configurations. Then, we used the idea of Cencek and Kutzelnigg \cite{CeKu97} 
for rescaling the centers upon making a small displacement of the nuclear configurations
(also used by Pavanello and Adamowicz for the computation of triatomic molecules
\cite{PaAd12}).
By making sufficiently small nuclear displacements, one obtains a very good starting
(non-linear) parameter set for the fECGs, which can be refined at a moderate computational cost. 
For the numerical examples shown later (Section~\ref{ch:numres} and \papertwo)
we used a 0.1~bohr step size for the intermolecular distance and 
carried out 1-2 full refinement cycle was sufficient to maintain 
the accuracy of the electronic energies along the potential energy curve.
%

\subsection{Diagonal Born--Oppenheimer correction}
Since the pragmatic approach of Handy and co-workers \cite{HaYaSc86,IoAmHa96,HaLe96}, 
it is possible to compute
the diagonal Born--Oppenheimer correction (DBOC) using the simple and general
expression in laboratory-fixed Cartesian coordinates:
\begin{align}
  U_{ia}(\mx{R})
  &=
  \frac{1}{2}
  \left\langle
    \pd{\psi}{R_{ia}} {\Big|}
    \pd{\psi}{R_{ia}}
  \right\rangle_\el 
  \label{eq:lfccdboc}
\end{align}
We compute the wave function derivatives numerically using the rescaling
idea of Cencek and Kutzelnigg \cite{CeKu97}
(since the displacements were very small, on the order of $10^{-4}$~bohr,
the full refinement of the basis set was not necessary):
\begin{align}
  \pd{\psi}{R_{ia}}(\mx{R})
  \approx
  \frac{%
    \psi(\bos{R}+\frac{1}{2}\bos{\Delta}_{ia})
    -
    \psi(\bos{R}-\frac{1}{2}\bos{\Delta}_{ia}) 
  }{%
    \Delta_{ia}
  }
\end{align}
with the $\bos{\Delta}_{ia}$ displacement vector (of the atomic nuclei), which
labels a $\Delta_{ia}$ displacement along the $ia$ degree of freedom.
Then, Eq.~(\ref{eq:lfccdboc}) becomes
\begin{align}
  U_{ia}(\mx{R}) 
  &\approx 
    \frac{(1-S^{(\pm)}_{ia}(\mx{R}))}{(\Delta_{ia})^2} ,
\end{align}
where we exploited that the electronic wave function is real and normalized
and 
\begin{align}
  S_{ia}^{(\pm)}(\mx{R})
  =
  \langle
    \psi(\mx{R}+\frac{1}{2}\bos{\Delta}_{ia})  |
    \psi(\mx{R}-\frac{1}{2}\bos{\Delta}_{ia}) 
  \rangle_\el  .
\end{align}

\subsection{Mass-correction tensor}
The mass-correction tensor, Eq.~(\ref{eq:masscorrtensor}),
for an isolated electronic state $(E,\psi)$is written
in Cartesian coordinates as (the bar over $\mmx$ indicates that the electronic-structure 
computations are carried out for a selected embedding of the nuclei):
\begin{align}
  \bar{\mmx}_{ia,jb}(\mx{R})
  &=
  2\langle%
    \partial_{R_{ia}}\psi|
    (\hel-\eel)^{-1}(1-\hat{P})
    \partial_{R_{jb}}\psi
  \rangle_\el . 
\end{align}
In order to calculate the effect of the resolvent, we
introduce an auxiliary basis set 
$\lbrace f_n,n=1,2,\ldots,\naux \rbrace$:
\begin{align}
  \bar{\mmx}_{ia,jb}(\mx{R})
  &=
  2\sum_{n=1}^{\naux}
  \sum_{m=1}^{\naux}
  \langle %
    \partial_{R_{ia}}\psi| f_n 
  \rangle_\el
  \left(\bos{F}^{-1}\right)_{nm}
  \langle %
    f_m |\partial_{R_{jb}}\psi
  \rangle_\el 
  \label{eq:Amxauxbas}
\end{align}
with 
\begin{align}
  (\bos{F})_{nm}
  =
  \langle 
    f_n |
    (\hel - \eel) (1-\hat{P})
    | f_m
  \rangle_\el.
  \label{eq:Fmx}
\end{align}
Representation of the resolvent by the direct summation over excited electronic
states is impractical, because it would require to compute (and converge) a very large number 
of electronic states to tightly converge
the resolvent (and the mass-correction tensor). 
Instead, we ensure the convergence of the mass matrix elements by enlarging the 
auxiliary basis set similarly to Refs.~\cite{PaKo09,prx17}.
As for the DBOC, the wave function derivatives are computed by finite differences
using the fECG rescaling idea \cite{CeKu97}: 
\begin{align}
  \langle %
    \partial_{R_{ia}}\psi| f_n 
  \rangle_\el
  &\approx
  \frac{1}{\Delta_{ia}}
  \left(%
    T^{(+)}_{ia,n}
    -
    T^{(-)}_{ia,n}
  \right) ,
  \label{eq:findiff}
\end{align}
where we have introduced the short notation
\begin{align}
  T^{(\pm)}_{ia,n}
  &=
  \langle%
    \psi(\mx{R}\pm \frac{1}{2} \bos{\Delta}_{ia}) | f_n
  \rangle_\el
  \nonumber \\
  &=
  \sum_{k=1}^{N^{(\pm)}_{ia}}
    (\bos{c}_{ia}^{(\pm)})_k
    \langle %
      (g_{ia}^{(\pm)})_k|f_n
    \rangle_\el .
  \label{eq:tmx}
\end{align}
In Eq.~(\ref{eq:tmx}), $(g_{ia}^{(\pm)})_k$ is the $k$th
fECG basis function obtained for 
the $\mx{R}\pm\bos{\Delta}_{ia}$ nuclear geometry
and $(\bos{c}_{ia}^{(\pm)})_k$ is the corresponding linear combination
coefficient of the electronic state $\psi(\mx{r};\mx{R}\pm\bos{\Delta}_{ia})$ 
resulting from the variational solution
of the electronic Schrödinger equation in the $(g_{ia}^{(\pm)})_k$ 
basis set ($k=1,2,\ldots,N^{(\pm)}_{ia}$).

\paragraph{Evaluation of the resolvent}
%
Instead of directly constructing the $\mx{F}^{-1}$ matrix, 
it is computationally more accurate and stable (see also Ref.~\cite{prx17}) to
consider
\begin{align}
  (\mx{F}\ \mx{x}_{j_b})_m
  = 
  \langle %
    f_m |\partial_{R_{jb}}\psi
  \rangle_\el 
  \label{eq:Fxb}
\end{align}
solve this system of linear equations for $\mx{x}_{j_b}$, 
and then evaluate Eq.~(\ref{eq:Amxauxbas}) as the sum of vector products.
Furthermore, we consider $\mx{F}(\eel+\varepsilon)$ 
instead of $\mx{F}(\eel)$ with a small $\varepsilon$ real or imaginary value 
in order to avoid numerical instabilities due to having either explicitly or
implicitly the inverse of a singular matrix 
(note that $E$ is an eigenvalue of $\hat{H}$ in Eq~(\ref{eq:Fmx})). 
The optimal value of $\varepsilon$ is determined in a series of computations 
by maximizing both numerical stability and accuracy.

\paragraph{Auxiliary basis set}
In numerical applications, it is important to ensure that 
Eq.~(\ref{eq:Amxauxbas}) is converged with respect
to the enlargement of the auxiliary basis set. A reasonable starting point 
is to use the basis set optimized for the solution of the electronic Schr\"odinger
equation. This choice often gives an accurate estimate 
(at least within a 1-2~\% of the exact value), but usually additional
functions are necessary to obtain really accurate mass-correction terms.
Pachucki and Komasa optimized the auxiliary basis set by using the variational
property of the rotational and vibrational mass-correction terms arising in
the curvilinear expression of the mass-correction tensor for diatomic
molecules \cite{PaKo09}.
In Ref.~\cite{prx17} a plane-wave expansion was used both for the electronic-state 
representation and also for the auxiliary basis set.

We do not have any direct and general optimization strategy to
build an optimal and converged auxiliary basis set of fECGs for general
polyatomic and polyelectronic molecules. Instead, we simply enlarge
the auxiliary basis set until convergence was achieved by using
the electronic basis sets optimized at neighbohring nuclear configurations.
As a practical handle, we have noticed that 
if the resolution of identity was poorly represented by the auxiliary
basis functions, the resolvent was also inaccurate.
The resolution of identity in a non-orthogonal set of functions is
\begin{align}
  \hat{I}
  &=
  \sum_{i=1}^{\naux}
  \sum_{j=1}^{\naux}
    \left. | f_i \rangle\ 
    \cdot 
    \langle f_i| \hat{I} | f_j \rangle_\el^{-1}
    \cdot
    \langle f_j | \right.
  =
  \sum_{i=1}^{\naux}
  \sum_{j=1}^{\naux}
    \left. | f_i \rangle\ 
    \cdot S_{ij}^{-1} \cdot
    \langle f_j | \right.
\end{align}
and we evaluate the derivative overlaps also by inserting the auxiliary basis set:
\begin{align}
 \langle \partial_k \Psi | \partial_k \Psi \rangle_\el
 \approx
 \langle \partial_k \Psi | f_i \rangle_\el\ 
 \cdot S_{ij}^{-1} \cdot
 \langle f_j | \partial_k \Psi \rangle_\el
 \quad\text{with}\quad S_{ij} = \langle f_i|f_j \rangle_\el .
 \label{eq:dbocaux}
\end{align}
The accuracy of Eq.~(\ref{eq:dbocaux}) is not a sufficient condition 
for the convergence of Eq.~(\ref{eq:Amxauxbas}), it is merely a useful indicator
for the reliability of the results.

\paragraph{Frame transformation}
It is important to notice that the mass-tensor elements are evaluated
with a certain frame definition of the nuclear geometry.  This choice of 
the body-fixed frame in the electronic-structure computations is described 
by the $\mx{O}$ rotation matrix which connects this body-fixed frame and the laboratory-fixed
frame. 
The body-fixed frame used to compute the mass-correction tensor may be different
from the body-fixed frame used in the rovibrational computations.
The general curvilinear form of the second-order non-adiabatic
kinetic energy operator, Eqs.~(\ref{eq:divMgradGtilde})--(\ref{eq:smxdef}),
includes the corresponding frame definitions.

\clearpage

\section{Applications for homonuclear diatomics \label{ch:applications}}

\subsection{Evaluation of the general curvilinear kinetic energy expressions for homonuclear diatomic molecules}
The quantum mechanical motion of atomic nuclei in di- and poly-atomic molecules is efficiently described if the 
laboratory-frame (LF) Cartesian coordinates, $R_1,\ldots,R_{3N}$, are replaced with internal coordinates,
orientation angles, 
and the Cartesian coordinates of the nuclear center of mass (NCM), $\mx{R}_\text{NCM}$. 

The general curvilinear kinetic-energy operator expressions developed in Section~\ref{ch:coortrfo}
can be directly implemented in the GENIUSH program and the derived effective mass-matrix expressions 
can be evaluated at grid points similarly to the mass-weighted metric tensor as
it has been implemented for the constant mass case in Ref.~\cite{MaCzCs09}.
In this section, we shall go through the calculations with the coordinate-dependent 
mass-correction tensor step by step and derive
the expressions explicitly 
in order to better understand the formalism and to highlight the formal and numerical 
properties of the derived expressions for the simple case of homonuclear diatomic molecules.

\paragraph{Preparatory calculations with spherical polar coordinates}
The transformation from the $(r^1,r^2,r^3)$ ``flat coordinates'' to 
spherical polar coordinates, 
$\rho\in[0,\infty)$, $\theta\in(-\pi,\pi)$, and $\phi\in[0,2\pi)$, is defined by
\begin{align}
  \left(%
    \begin{array}{@{}c@{}}
      r^1 \\
      r^2 \\
      r^3 \\
    \end{array}
  \right)
  =  
  \left(%
    \begin{array}{@{}c@{}}
      \rho \sin\theta \cos\phi \\
      \rho \sin\theta \sin\phi \\
      \rho \cos\theta \\
    \end{array}
  \right)  .
\end{align}
the corresponding Jacobian matrix is
\footnote{In matrices, we shall use ``.'' for ``0'' to enhance readability in matrix expressions.}:
\begin{align}
  \jmxsph
  =
  \left(%
    \begin{array}{@{}ccc@{}}
      \sin\theta \cos\phi &  \rho\cos\theta\cos\phi & -\rho\sin\theta\sin\phi \\
      \sin\theta \sin\phi &  \rho\cos\theta\sin\phi &  \rho\sin\theta\cos\phi \\
      \cos\theta          & -\rho\sin\theta         &  . \\
    \end{array}
  \right). 
  \label{eq:jmxsph}
\end{align}
The covariant metric tensor, the elements of which are calculated as
$g_{\mu\nu}=\delta_{ij}\pd{r^i}{\xi^\mu} \pd{r^i}{\xi^\nu}$, is: 
\begin{align}
  g_{\mu\nu} 
  =
  \left(%
    \begin{array}{@{}ccc@{}}
      1 & . & . \\
      . & \rho^2 & . \\      
      . & . & \rho^2\sin^2\theta \\  
    \end{array}
  \right),
\end{align}
and it is inverted to obtain, the contravariant metric tensor:
\begin{align}
  g^{\mu\nu} 
  =
  \left(%
    \begin{array}{@{}ccc@{}}
      1 & . & . \\
      . & \frac{1}{\rho^2} & . \\      
      . & . & \frac{1}{\rho^2\sin^2\theta} \\  
    \end{array}
  \right).
\end{align}
The Jacobi determinant for this coordinate transformation reads as:
\begin{align}
  \tlg^{-1/2}
  =
  \left[\det(g^{\mu\nu})\right]^{-1/2}
  =
  \det(\jmxsph)
  =
  \rho^2\sin\theta.
\end{align}
The electronic structure computations and the evaluation 
of the mass-correction tensor is carried out in a frame selected for the atomic nuclei.
For homonuclear diatomic molecules 
it is a natural choice to position the nuclei symmetrically with respect to the origin along
the $z$ axis. 
The rotation matrix from this ``$z$-axis embedding''
to the laboratory-fixed frame (expressed with the $\theta$ and $\phi$ spherical angles) is:
\begin{align}
  \omxsph
  &=
  \left(%
    \begin{array}{@{}ccc@{}}
      \cos\phi & -\sin\phi & . \\
      \sin\phi & \cos\phi & . \\
      . & . & 1 \\
    \end{array}
  \right)
  \cdot
  \left(%
    \begin{array}{@{}ccc@{}}
       \cos\theta & . & \sin\theta \\
      . & 1 & . \\
      -\sin\theta & . & \cos\theta \\
    \end{array}
  \right)
  \nonumber \\
  &=
  \left(%
    \begin{array}{@{}c@{\ \ \ }c@{\ \ \ }c@{}}
       \cos\theta\cos\phi & -\sin\phi & \sin\theta\cos\phi \\
      \cos\theta\sin\phi & \cos\phi & \sin\theta\sin\phi \\
      -\sin\theta & . & \cos\theta \\
    \end{array}
  \right) .
  \label{eq:omxsph}
\end{align}
Thus, the $\mx{S}$ transformation matrix defined in Eq.~(\ref{eq:smxdef})
for these curvilinear coordinates and
body-fixed frame is:
\begin{align}
  \smxsph
  =
  \jmxsph\tr
  \omxsph
  =
  \left(%
    \begin{array}{@{}ccc@{}}
      . & . & 1 \\ 
      r & . & . \\
      . & r\sin\theta & . \\
    \end{array}
  \right).
  \label{eq:Smxsph}
\end{align}
Note also the relationship between the covariant metric tensor and 
the $\smxsph$ transformation matrix:
\begin{align}
  \mx{g}^{-1}\smxsph\smxsph\tr
  =
  \smxsph\smxsph\tr\mx{g}^{-1}
  =\mx{I}_3 .
\end{align}

\subsection{Transformation to spherical polar and center-of-mass Cartesian coordinates}
For diatomic molecules, the six laboratory-fixed Cartesian coordinates 
are replaced with three NCM Cartesian coordinates 
and the three spherical polar coordinates. 
So, the $(R^1,\ldots,R^6)\rightarrow(\xi^1,\ldots,\xi^6)$ transformation from the $R^i$ 
``flat coordinates'' to the $\xi^\mu$ 
new coordinates is 
\begin{align}
  \left(%
    \begin{array}{@{}c@{}}
      R^1 \\
      R^2 \\
      R^3 \\
      \cline{1-1}
      R^4 \\      
      R^5 \\
      R^6 \\      
    \end{array}
  \right)
  =
  \left(%
    \begin{array}{@{}c@{}}
      -\frac{1}{2}r^1 + R^1_{\text{NCM}} \\
      -\frac{1}{2}r^2 + R^2_{\text{NCM}} \\
      -\frac{1}{2}r^3 + R^3_{\text{NCM}} \\
      \cline{1-1}    
      \frac{1}{2}r^1 + R^1_{\text{NCM}} \\
      \frac{1}{2}r^2 + R^2_{\text{NCM}} \\
      \frac{1}{2}r^3 + R^3_{\text{NCM}} \\
    \end{array}
  \right)
  =
  \left(%
    \begin{array}{@{}c@{}}
      -\frac{1}{2}\rho \sin\theta \cos\phi + R^1_{\text{NCM}} \\
      -\frac{1}{2}\rho \sin\theta \sin\phi + R^2_{\text{NCM}} \\
      -\frac{1}{2}\rho \cos\theta + R^3_{\text{NCM}} \\
      \cline{1-1}    
       \frac{1}{2}\rho \sin\theta \cos\phi + R^1_{\text{NCM}} \\
       \frac{1}{2}\rho \sin\theta \sin\phi + R^2_{\text{NCM}} \\
       \frac{1}{2}\rho \cos\theta + R^3_{\text{NCM}} \\
    \end{array}
  \right)  
\end{align}
where the $\mx{r}$ internuclear displacement vector is written in terms of spherical polar coordinates 
and $\mx{R}_{\text{NCM}}$ collects the center of mass coordinates of the atomic nuclei 
(calculated with the nuclear masses).
The Jacobian matrix of this transformation 
is
\begin{align}
  \jmxdiat
  =
  \ '' \pd{R^i}{\xi^\mu} ''
  =
  \left(%
    \begin{array}{@{}c@{\ \ \ \ \ \ }c@{\ \ \ \ \ \ }c  c | c@{\ \ \ \ }c@{\ \ \ \ }c@{}}
      & & &                                         &    1 & . & . \\          
      \multicolumn{3}{c}{-\frac{1}{2}\jmxsph} &&    . & 1 & . \\
      & & &                                         &    . & . & 1 \\            
      \cline{1-7}
      & & &                                         &    1 & . & . \\          
      \multicolumn{3}{c}{\frac{1}{2}\jmxsph} &&     . & 1 & . \\
      & & &                                         &    . & . & 1 \\      
    \end{array}
  \right) ,
\end{align}
where $\jmxsph$ is the Jacobian matrix corresponding to the spherical polar coordinates given 
in Eq.~(\ref{eq:jmxsph}).
The covariant metric tensor is
\begin{align}
  g_{\mu\nu} 
  =
  \ ''  \delta_{ij}\pd{R^i}{\xi^\mu} \pd{R^i}{\xi^\nu} ''
  = 
  \left(%
    \begin{array}{@{}ccc | ccc@{}}
      \frac{1}{2} & . & . &    . & . & . \\
      . & \frac{1}{2}\rho^2 & . &    . & . & . \\      
      . & . & \frac{1}{2}\rho^2\sin^2\theta &    . & . & . \\      
      \cline{1-6}
      . & . & . &   2 & . & . \\
      . & . & . &   . & 2 & . \\
      . & . & . &   . & . & 2 \\      
    \end{array}
  \right) ,
  \label{eq:covargdiat}
\end{align}
and the Jacobi determinant reads as
\begin{align}
  \tlg^{1/2} = \rho^2\sin\theta.
  \label{eq:Jacdet6d}
\end{align}
The rotation of the 6-dimensional position vectors expressed
in the $z$-axis embedding, which is also the body-fixed frame 
which we use in electronic structure theory to compute the mass-correction tensor, 
and in the laboratory frame is the direct product of the rotation matrix given in 
Eq.~(\ref{eq:omxsph}) with the $2\times 2$ unit matrix:
\begin{align}
  \omxdiat 
  = 
  \omxsph \otimes \mx{I}_2 ,
\end{align}
and thus, 
the $\mx{S}$ transformation matrix defined in Eq.~(\ref{eq:smxdef})
for this 6-dimensional coordinate transformation is
\begin{align}
  \smxdiat
  &=
  \jmxdiat\tr\omxdiat 
  =
  \left(%
    \begin{array}{@{}c|c@{}}
      -\frac{1}{2}\smxsph & \frac{1}{2}\smxsph \\
      \cline{1-2}
       \omxsph & \omxsph \\
    \end{array}
  \right),
  \label{eq:smx6d}
\end{align}
where $\smxsph$ was given in Eq.~(\ref{eq:Smxsph}).

\subsection{Calculation of the BO and non-adiabatic kinetic energy operators in curvilinear coordinates}
For a start, let us consider (and reproduce) the kinetic energy operator for 
the constant-mass case, \emph{i.e.,} for a constant, diagonal mass matrix 
($1/\massnuc \mx{I}_6$ with the $\mx{I}_6$ $6\times 6$ unit matrix):
\begin{align}
  &\hat{T}^{(0)}
  \nonumber\\
  &=
  -\frac{1}{2\massnuc}
  \div \bos{I}_6 \grad 
  \nonumber \\
  &=
  -\frac{1}{2\massnuc}  
  \tlg^{-1/2}
  \partial_\mu
  \tlg^{1/2}
  g^{\mu\alpha}
  (\mx{S}\mx{I}_6\mx{S}\tr)_{\alpha\beta}
  g^{\beta\nu}
  \partial_\nu 
  \nonumber \\
  &=
  -\frac{1}{2\massnuc}  
  \tlg^{-1/2}
  \partial_\mu
  \tlg^{1/2}
  g^{\mu\nu}
  \partial_\nu 
  \nonumber \\
  &=
  -\frac{1}{2\massnuc}  
  \frac{1}{\rho^2\sin\theta}
  \left(%
    \partial_\rho,\partial_\theta,\partial_\phi,\partial_{R^1_\text{NCM}},
    \partial_{R^2_\text{NCM}},\partial_{R^3_\text{NCM}}
  \right)
  \rho^2\sin\theta
  \left(%
    \begin{array}{@{}ccc ccc@{}}
      2 & . & . &   . & . & . \\
      . & \frac{2}{\rho^2} & . &   . & . & . \\
      . & . & \frac{2}{(\rho\sin\theta)^2} &   . & . & . \\
      . & . & . &   \frac{1}{2} & . & . \\
      . & . & . &   . & \frac{1}{2} & . \\
      . & . & . &   . & . & \frac{1}{2} \\
    \end{array}
  \right)
  \left(%
    \begin{array}{@{}c@{}}
      \partial_\rho \\
      \partial_\theta \\
      \partial_\phi \\
      \partial_{R^1_\text{NCM}} \\
      \partial_{R^2_\text{NCM}} \\
      \partial_{R^3_\text{NCM}} \\
    \end{array}
  \right)
  \nonumber \\
  &=
  -
  \frac{1}{2}
  \frac{1}{\rho^2\sin\theta}
  (\partial_\rho,\partial_\theta,\partial_\phi)
  \rho^2\sin\theta
  \left(%
    \begin{array}{@{}ccc@{}}
      \frac{2}{\massnuc}  & .  & . \\
      .  &  \frac{2}{\massnuc}\frac{1}{\rho^2}  & . \\
      .  & .  & \frac{2}{\massnuc}\frac{1}{\rho^2\sin^2\theta}  \\       
    \end{array}
  \right)  
  \left(%
    \begin{array}{@{}c@{}}
      \partial_\rho \\
      \partial_\theta \\
      \partial_\phi \\
    \end{array}
  \right)
  +\hat{T}^{(0)}_\text{NCM},
  \label{eq:BOkin} 
\end{align}
where $\partial_\mu=\partial/\partial \xi^\mu$ and the kinetic energy operator of 
the nuclear center of mass was introduced as:
\begin{align}
   \hat{T}^{(0)}_\text{NCM} 
  =-
  \frac{1}{2}  
  \frac{1}{2\massnuc}  
  \left(%
     \partial^2_{R^1_{\text{NCM}}}
    +\partial^2_{R^2_{\text{NCM}}}
    +\partial^2_{R^3_{\text{NCM}}}   
  \right) .
  \label{eq:BOkinNCM}
\end{align}
It is important to remember that $\hat{T}^{(0)}_\text{NCM}$ 
is separable from the rovibrational kinetic energy operator.

In Eq.~(\ref{eq:BOkin}) we have reproduced the well-known diatomic 
rotation-vibration kinetic energy expression, which is the first term including 
the inverse of the reduced mass of the two identical nuclei, 
$1/\mu = 1/\massnuc + 1/\massnuc = 2/\massnuc$,
and the translational kinetic energy of the nuclear center of mass, 
which is the second term including 
the inverse of the total nuclear mass, $1/M = 1/(2\massnuc)$.

Next, let us calculate the curvilinear kinetic energy operator including 
the mass-correction tensor 
for the present example. 
In addition to the general expressions, to better highlight how the formalism is used, 
we also give the numerical results 
for the H$_2^+$ molecular ion for $\rho=3$~bohr internuclear separation. 

The general form of the mass-correction tensor for a homonuclear diatomic molecule 
in the $z$-axis embedding can be deduced from simple symmetry arguments
($A,B,a,b\in\mathbb{R}^+$, further conditions apply, \emph{vide infra}):
\begin{align}
  \bar{\mmx}_{ij}
   &=
  \left(%
    \begin{array}{@{}ccc ccc @{}}
      A &  . &  . & a & . & . \\
       . & A & . & . & a & . \\
       . & . & B & . & . & b \\
      a & . & . & A & . & . \\
       . & a & . & . & A & . \\
       . & . & b & . & . & B \\
   \end{array}
 \right) 
 \label{eq:amxdiat}
 \quad \text{$\lbrace\rho\in\mathbb{R}^+,\ z\text{-axis embedding}\rbrace$}, 
\end{align}
which takes the numerical values 
\begin{align}
  \bar{\mmx}_{ij}
  &=
  \left(%
    \begin{array}{@{}ccc ccc @{}}
      0.322 &  . &  . & 0.178 & . & . \\
       . & 0.322 & . & . & 0.178 & . \\
       . & . & 0.462 & . & . & 0.038 \\
      0.178 & . & . & 0.322 & . & . \\
       . & 0.178 & . & . & 0.322 & . \\
       . & . & 0.038 & . & . & 0.462 \\
   \end{array}
 \right)
 \quad \text{$\lbrace \text{for}\ \rho=3\ \text{bohr}\rbrace$.}
\end{align}
The curvilinear expression for $\mmx_{\alpha\beta}$ is 
obtained by using the $\mx{S}$ 
transformation matrix given in Eq.~(\ref{eq:smx6d})
of the present coordinate transformation:
%
\begin{align}
  \mmx_{\alpha\beta}
  &=
  (\mx{S}\bar{\mx{\mmx}}\mx{S}\tr)_{\alpha\beta}
  \nonumber \\
  &=
  \left(%
    \begin{array}{@{}ccc | ccc @{}}
      \frac{B-b}{2} &  . &  . & . & . & . \\
       . & \frac{A-a}{2}\rho^2 & . & . & . & . \\
       . & . & \frac{A-a}{2}\rho^2\sin^2\theta & . & . & . \\
       \cline{1-6}
       . & . & . & 2(B+b) & . & . \\
       . & . & . & . & 2(A+a) & . \\
       . & . & . & . & . & 2(A+a) \\
   \end{array}
 \right)
 \nonumber \\ 
 &\quad\text{$\lbrace \rho\in\mathbb{R}^+, (\rho,\theta,\phi)\rbrace$,} 
 \label{eq:amxcurvdiat} \\
 \text{which equals to} \nonumber\\
  &=
  \left(%
    \begin{array}{@{}ccc | ccc @{}}
      0.212 &  . &  . & . & . & . \\
       . & 0.072\rho^2 &  . & . & . & . \\
       . & . & 0.072\rho^2\sin^2\theta & . & . & . \\
       \cline{1-6}
       . &  . &  . & 1. & .  & . \\
       . &  . &  . & .  & 1. & . \\
       . &  . &  . & .  & .  & 1. \\       
     \end{array}
 \right)
 \quad \text{$\lbrace \text{for}\ \rho=3\ \text{bohr}\rbrace$.}
\end{align}
In the next step, we multiply $\mmx_{\alpha\beta}$ with 
the contravariant metric tensor,
and obtain the following simple expression
\begin{align}
  \mmx^\mu\,_\beta\ 
  &=
  g^{\mu\alpha}\mmx_{\alpha\beta}
  =
  g^{\mu\alpha}(\mx{S}\bar{\mx{\mmx}}\mx{S}\tr)_{\alpha\beta}
  \nonumber \\
  &=
  \left(%
    \begin{array}{@{}ccc | ccc @{}}
      B-b & . &  . & . & . & . \\
       . & A-a & . & . & . & . \\
       . & . & A-a & . & . & . \\
       \cline{1-6}
       . & . & . & B+b & . & . \\
       . & . & . & . & A+a & . \\
       . & . & . & . & . & A+a \\
   \end{array}
 \right)
 \quad\text{$\lbrace \rho\in\mathbb{R}^+,\ (\rho,\theta,\phi)\rbrace$}
 \label{eq:tamxcurvdiat}\\
 \nonumber \\
 &=
  \left(%
    \begin{array}{@{}ccc | ccc @{}}
      0.424 & . &  . & . & . & . \\
       . & 0.144 & . & . & . & . \\
       . & . & 0.144 & . & . & . \\
       \cline{1-6}
       . & . & . & 0.5 & . & . \\
       . & . & . & . & 0.5 & . \\
       . & . & . & . & . & 0.5 \\
   \end{array}
 \right)
 \quad \text{$\lbrace \text{for}\ \rho=3\ \text{bohr}\rbrace$.}
\end{align}
Furthermore, by multiplying this matrix with $g^{\beta\nu}$ from the right, we
get
\begin{align}
  \mmx^{\mu\nu} 
  &=\mmx^\mu\,_\beta\ 
  g^{\beta\nu} \\  
  &=
  \rho^2\sin\theta
  \left(%
    \begin{array}{@{}ccc | ccc @{}}
      2(B-b)  &  . &  . & . & . & . \\
       . & 2(A-a) \frac{1}{\rho^2} &  . & . & . & . \\
       . & . & 2(A-a) \frac{1}{\rho^2\sin^2\theta} & . & . & . \\
       \cline{1-6}
       . &  . &  . & \frac{B+b}{2} & .  & . \\
       . &  . &  . & .  & \frac{A+a}{2} & . \\
       . &  . &  . & .  & .  & \frac{A+a}{2} \\       
     \end{array}
  \right)  
  \quad\text{$\lbrace \rho\in\mathbb{R}^+,\ (\rho,\theta,\phi)\rbrace$}
  \label{eq:mmmxcurv}
  \\ 
  &=
   \rho^2\sin\theta
   \left(%
     \begin{array}{@{}ccc | ccc @{}}
       0.848 &  . &  . & . & . & . \\
        . & 0.288\frac{1}{\rho^2} &  . & . & . & . \\
        . & . & 0.288\frac{1}{\rho^2\sin^2\theta} & . & . & . \\
        \cline{1-6}
        . &  . &  . & 0.25 & .  & . \\
        . &  . &  . & .  & 0.25 & . \\
        . &  . &  . & .  & .  & 0.25 \\       
      \end{array}
  \right)  
  \quad\text{$\lbrace \rho=3,\ (\rho,\theta,\phi)\rbrace$,} 
  %
\end{align}
and we write down the 
effective $\tilde{G}^{\mu\nu}$ matrix defined in Eq.~(\ref{eq:geff}) as:
\begin{align}
  &\tilde{G}^{\mu\nu}
  =
  \frac{1}{{\massnuc}}
  g^{\mu\alpha}
  (%
    \mx{S}
    \mx{I}_6
    \mx{S}\tr
  )_{\alpha\beta}  
  g^{\beta\nu}
  -
  \frac{1}{{\massnuc}^2}
  g^{\mu\alpha}
  (%
    \mx{S}
    \bar{\mmx}
    \mx{S}\tr
  )_{\alpha\beta}  
  g^{\beta\nu}
  \noindent \\
  &=
  \left[%
    \delta^\mu\,_\beta
    -
    \frac{1}{{\massnuc}}  
    \mmx^\mu\,_\beta\ 
  \right]
  \frac{1}{{\massnuc}}  
  g^{\beta\nu}
  \\  
  &=
  {\tiny
  \left(%
    \begin{array}{@{}c@{}c@{}c | c@{}c@{}c @{}}
      \frac{2}{\massnuc}\left[1-\frac{\mmx^\rho\,_\rho}{\massnuc}\right] &  . &  . & . & . & . \\
       . & \frac{2}{\massnuc}\left[1-\frac{\mmx^\theta\,_\theta}{\massnuc}\right] \frac{1}{\rho^2} &  . & . & . & . \\
       . & . & \frac{2}{\massnuc}\left[1-\frac{\mmx^\phi\,_\phi}{{\massnuc}}\right]\frac{1}{\rho^2\sin^2\theta} & . & . & . \\
       \cline{1-6}
       . &  . &  . & \frac{1}{2\massnuc}\left[1-\frac{\mmx^{R_1}\,_{R_1}}{{\massnuc}}\right] & . & . \\
       . &  . &  . &  . & \frac{1}{2\massnuc}\left[1-\frac{\mmx^{R_2}\,_{R_2}}{{\massnuc}}\right] & . \\
       . &  . &  . &  . & . & \frac{1}{2\massnuc}\left[1-\frac{\mmx^{R_3}\,_{R_3}}{{\massnuc}}\right] \\
     \end{array}
 \right)   
 }\nonumber 
 \\  
  &=
  {\tiny
  \left(%
    \begin{array}{@{}c@{}c@{}c | c@{}c@{}c @{}}
      \frac{2}{\massnuc}\left[1-\frac{B-b}{\massnuc}\right] &  . &  . & . & . & . \\
       . & \frac{2}{\massnuc}\left[1-\frac{A-a}{\massnuc}\right] \frac{1}{\rho^2} &  . & . & . & . \\
       . & . & \frac{2}{\massnuc}\left[1-\frac{A-a}{{\massnuc}}\right]\frac{1}{\rho^2\sin^2\theta} & . & . & . \\
       \cline{1-6}
       . &  . &  . & \frac{1}{2\massnuc}\left[1-\frac{B+b}{{\massnuc}}\right] & . & . \\
       . &  . &  . &  . & \frac{1}{2\massnuc}\left[1-\frac{A+a}{{\massnuc}}\right] & . \\
       . &  . &  . &  . & . & \frac{1}{2\massnuc}\left[1-\frac{A+a}{{\massnuc}}\right] \\
     \end{array}
 \right)   
 } 
 \label{eq:diatGeff}
 \\ 
 &\quad\text{for $\rho\in\mathbb{R}^+$ (spherical polar coordinates, $z$-axis embedding).}
 \nonumber
\end{align}
By inserting the mass-correction values for $\rho=3$~bohr we have
\begin{align}
  &=
  \left(%
    \begin{array}{@{}c@{}c@{}c | c@{}c@{}c @{}}
      \frac{2}{\massnuc}\left[1-\frac{0.424}{\massnuc}\right] &  . &  . & . & . & . \\
       . & \frac{2}{\massnuc}\left[1-\frac{0.144}{\massnuc}\right] \frac{1}{\rho^2} &  . & . & . & . \\
       . & . & \frac{2}{\massnuc}\left[1-\frac{0.144}{{\massnuc}}\right]\frac{1}{\rho^2\sin^2\theta} & . & . & . \\
       \cline{1-6}
       . &  . &  . & \quad\quad\quad\quad  & \quad\quad  &   \\
       . &  . &  . & \multicolumn{3}{c}{\frac{1}{2\massnuc}\left[1-\frac{1}{2{\massnuc}}\right]\mx{I}_3} \\
       . &  . &  . & \quad\quad\quad\quad  & \quad\quad &  \\       
     \end{array}
 \right)   
 \nonumber
 \\ 
 &\quad\quad \text{for H$_2^+$ with $\rho=3$~bohr (spherical polar coordinates, $z$-axis embedding).}
\end{align}
Using this effective $\tilde{G}^{\mu\nu}$ matrix and exploiting that
its off-diagonal elements are all zero, we
write the second-order nonadiabatic kinetic energy
operator for a diatomic molecule 
(for comparison see the constant-mass operator in Eq.~(\ref{eq:BOkin})--(\ref{eq:BOkinNCM}))
as
\begin{align}
  \hat{T}^{(2)}
  &=
  -\frac{1}{2}  
  \tlg^{-1/2}
  \partial_\mu
  \tlg^{1/2}
  \tilde{G}^{\mu\nu}
  \partial_\nu 
  \nonumber \\
  &=
  -\frac{1}{2}
  \frac{1}{\rho^2\sin\theta}
  \left(%
    \partial_\rho,\partial_\theta,\partial_\phi
  \right)
  \rho^2\sin\theta 
  \left(%
    \begin{array}{@{}ccc ccc@{}}
      \tilde{G}^{\rho\rho} & . & . \\
      . & \tilde{G}^{\theta\theta} & . \\
      . & . & \tilde{G}^{\phi\phi} \\
    \end{array}
  \right)
  \left(%
    \begin{array}{@{}c@{}}
      \partial_\rho \\
      \partial_\theta \\
      \partial_\phi \\
    \end{array}
  \right) 
  +
  \hat{T}^{(2)}_\text{NCM}.
  \label{eq:ttwodiatom}
\end{align}
The second-order non-adiabatic kinetic energy operator
for the nuclear center of mass motion is 
\begin{align}
  \hat{T}^{(2)}_\text{NCM}
  =-\frac{1}{2}  
  \tilde{G}^{\text{NCM}}
  \left(%
     \partial^2_{R_{\text{NCM},1}}
    +\partial^2_{R_{\text{NCM},2}}
    +\partial^2_{R_{\text{NCM},3}}    
  \right) 
  \label{eq:nonadkin}
\end{align}
and $\tilde{G}^{\text{NCM}}$ corresponds to the diagonal elements of the 
lower right block of the effective tensor given in Eq.~(\ref{eq:diatGeff}).

The resulting expressions have a couple of important and interesting properties.
By looking at the numerical values computed for H$_2^+$ 
(at $\rho=3$ and at other internuclear distance values) we may observe
that the non-zero, diagonal elements in the NCM block can be re-written into 
a physically meaningful form
using the $1-x\approx(1+x)^{-1}$ approximation, valid for small $x$ values:
\begin{align}
  \frac{1}{2\massnuc}\left[1-\frac{1}{2\massnuc}\right]
  \approx
  \frac{1}{2\massnuc [1+ 1/(2\massnuc) ]}
  =
  \frac{1}{2\massnuc + 1}
  =
  \frac{1}{2 \tilde{M}_\text{tot}}
  \label{eq:totmass}
\end{align}
where $\tilde{M}_\text{tot}$ equals the total mass of the H$_2^+$ molecular ion, two protons and one electron,
in atomic units ($\massel=1$).
Using the same approximation, we re-write the diagonal elements of the ``rovibrational''
block as
\begin{align}
  \frac{2}{\massnuc}
  \left[1-\frac{\delta\tilde{m}}{\massnuc}\right]
  \approx
  \frac{2}{\massnuc (1+\delta\tilde{m}/\massnuc)}
  =
  \frac{2}{\massnuc + \delta\tilde{m}}  
  =
  \frac{2}{\tmassnuc}  
  \label{eq:effmass}  
\end{align}
in which $\tmassnuc$ can be interpreted as an effective nuclear mass.
The ``effective vibrational mass'', corresponding to the $\rho$ degree of freedom, is thus
\begin{align}
  &\frac{2}{\massnuc}
  \left[%
    1-\frac{\mmx^\rho\,_\rho}{\massnuc}
  \right]
  \approx
  \frac{2}{\massnuc + \mmx^\rho\,_\rho}
  =\frac{2}{\tmassvib} ,
  \nonumber \\
  &\text{and}\quad
  \tmassvib = \massnuc + \mmx^\rho\,_\rho
  \label{eq:effvibmassgen} \\
  &\quad\quad 
  \text{$\lbrace \rho\in\mathbb{R}^+\ \text{bohr, spherical polar coordinates}\rbrace$,} \nonumber    
\end{align}
which has the value for $\rho=3$~bohr:
\begin{align}
  &\frac{2}{\massnuc}
  \left[%
    1-\frac{0.424}{\massnuc}
  \right]
  \approx
  \frac{2}{\massnuc + 0.424}
  =\frac{2}{\tmassvib}
  \nonumber \\
  &\text{and}\ \tmassvib = \massnuc+ 0.424 \label{eq:effvibmass}\\
  &\quad\quad
  \text{$\lbrace\rho=3\ \text{bohr, spherical polar coordinates}\rbrace$.}  
  \nonumber
\end{align}
The effective rotational mass, which is the same for the $\theta$ and $\phi$ degrees of freedom
($\mmx^\Omega\,_\Omega=\mmx^\theta\,_\theta=\mmx^\phi\,_\phi$),
is
\begin{align}
  &\frac{2}{\massnuc}
  \left[%
    1 - \frac{\mmx^\Omega\,_\Omega}{\massnuc}
  \right]
  \approx
  \frac{2}{\massnuc + \mmx^\Omega\,_\Omega}
  =\frac{2}{\tmassrot} ,
  \nonumber \\
  &\text{and}\quad
  \tmassrot = \massnuc + \mmx^\Omega\,_\Omega
  \label{eq:effrotmassgen}  \\
  &\quad\quad \text{$\lbrace \rho\in\mathbb{R}^+,\ \text{spherical polar coordinates}\rbrace$} \nonumber 
\end{align}
\begin{align}
  &\frac{2}{\massnuc}
  \left[%
    1 - \frac{0.144}{\massnuc}
  \right]
  \approx
  \frac{2}{\massnuc + 0.144}
  =\frac{2}{\tmassrot}
  \nonumber \\
  &\text{and}\ \tmassrot = \massnuc + 0.144   \label{eq:effrotmass}      \\
  &\quad
  \text{$\lbrace \rho=3\ \text{bohr, spherical polar coordinates}\rbrace$.}  \nonumber
\end{align}
%

There are two important, 
general properties of the second-order non-adiabatic kinetic energy operator,
$\ttwo$---which 
manifest themselves also in this example---, in relation with the transformation 
of the effective mass matrix
to translationally invariant (TI) coordinates and 
the Cartesian coordinates of the nuclear center of mass (NCM). 
A formal derivation for both properties was given in Ref.~\cite{prx17}.
First, the effective mass matrix is always block diagonal in 
a TI-NCM representation and there is not any coupling between
the block corresponding to translationally invariant coordinates
and the block of the nuclear-center-of-mass Cartesian coordinates.
Hence, the translational kinetic energy can always be separated 
from the rovibrational kinetic energy operator in $\ttwo$ 
(which is indeed a very important property). 
Second, the translational kinetic energy term in $\ttwo$,
can be rearranged (within the $1-1/\massnuc\approx (1+\massnuc)^{-1}$ approximation)
to a form in which the mass associated to the translational degrees of freedom 
is the total mass of the molecule (nuclei plus electrons). Hence,
in $\ttwo$ not only the rovibrational but also 
the translational kinetic energy 
gains a correction term,
which increases the total nuclear mass with the mass of the electrons.

In short, the overall translation 
remains exactly separable from the internal (rotational-vibrational)
degrees of freedom in the second-order kinetic-energy operator,
but has an effective mass, which is equal to the total mass of the molecule.
It is interesting to note that these general properties provided 
the starting point for the non-adiabatic theory
for second-order ``mass'' effects of Kutzelnigg \cite{Ku07}. This theory
was later used by Jaquet
and Kutzelnigg for diatomics \cite{JaKu08}, and by Jaquet and Khoma for 
modeling non-adiabatic effects in H$_3^+$  
\cite{JaKh17,JaKh18}.

\clearpage
\section{Computation of the mass correction functions and non-adiabatic corrections to 
the rovibrational energies\label{ch:numres}}
\noindent %
We consider the second-order non-adiabatic Hamiltonian 
with the effective $\tilde{G}^{\mu\nu}$ tensor including the mass-correction terms, 
see Eqs.~(\ref{eq:divMgrad})--(\ref{eq:geff}),
\begin{align}
  \left(%
    -\frac{1}{2\mnuc}\tlg^{-1/2}\partial_\mu\tlg^{1/2}\tilde{G}^{\mu\nu}\partial_\nu 
    + U + V
  \right)
    \phi = \enuc \phi
  \label{eq:rovibsolve}
\end{align}
and $U=\sum_{i=1}^N\sum_{a}U_{ia}/\mnuc$ is the diagonal Born--Oppenheimer correction to the BO potential energy, $V$.

For diatomic molecules, there is not any coupling term between the rotational
and vibrational degrees of freedom neither in the BO, $\tbo_\text{rv}=\tbo-\hat{T}^{(0)}_\text{NCM}$ in Eq.~(\ref{eq:BOkin}), 
nor in the second-order kinetic energy operator, $\ttwo_\text{rv}=\ttwo-\ttwo_\text{NCM}$ in (\ref{eq:ttwodiatom}).
Hence, the angular part of $\ttwo_\text{rv}$ can be integrated with the $Y_{JM}(\theta,\phi)$ 
spherical harmonic functions (similarly to the standard solution of diatomics with $\hat{T}_{\text{rv}}^{(0)}$), 
and we are left with the numerical solution of the radial equation:
\begin{align}
  &\left(%
    -\frac{1}{2}\frac{1}{\rho^2}
    \pd{}{\rho}
    \rho^2 
    \frac{2}{\massnuc}
    \left[%
      1-\frac{\mmx^\rho\,_\rho}{\massnuc}
    \right] 
    \pd{}{\rho}
  \right.
  \nonumber \\
  &+\left. 
    \frac{J(J+1)}{\rho^2}
    \frac{1}{\massnuc}
    \left[%
      1 - \frac{\mmx^\Omega\,_\Omega}{\massnuc}
    \right]
    + U(\rho) + V(\rho)
  \right)
  f_J(\rho) 
  =\enuc_J f_J(\rho) 
  \label{eq:diatrad1}
\end{align}
with the volume element $\rho^2\text{d}\rho$.
Instead of solving Eq.~(\ref{eq:diatrad1}), 
we proceed similarly to Pachucki and Komasa \cite{PaKo09} and use the operator identity
\begin{align}
 \frac{1}{\rho^2}\pd{}{\rho} \rho^2 X(\rho) \pd{}{\rho} f(\rho)
 =
 \frac{1}{\rho}\pd{}{\rho} X(\rho) \pd{}{\rho} \rho f(\rho)
 -\frac{1}{\rho} \pd{X}{\rho} f(\rho)
\end{align}
to obtain
\begin{align}
  &\left(%
    -\pd{}{\rho} 
    \frac{1}{\massnuc}
    \left[%
      1-\frac{\mmx^\rho\,_\rho}{\massnuc}
    \right] 
    \pd{}{\rho}
  \right.
  \nonumber \\
  &\left. 
   -\frac{1}{\rho}\frac{1}{\massnuc^2} \pd{\mmx^{\rho}\,_\rho}{\rho}
   +\frac{J(J+1)}{\rho^2}
    \frac{1}{\massnuc}
    \left[%
      1 - \frac{\mmx^\Omega\,_\Omega}{\massnuc}
    \right]
    +
    U(\rho) + V(\rho)
  \right)
  \phi_J(\rho) 
  =\enuc_J \phi_J(\rho)  
  \label{eq:diatrad2}
\end{align}
with the volume element $\text{d}\rho$.
To obtain rovibrational (in fact, rovibronic) energies and wave functions,
we solve Eq.~(\ref{eq:diatrad2}) using the discrete variable representation (DVR) and 
associated Laguerre polynomials, $L_n^{(\alpha)}$ with $\alpha=2$ for the
radial (vibrational) degree of freedom.
The DVR points are scaled to an $[R_\text{min},R_\text{max}]$ interval. 
The $n$ number of DVR points and functions, as well as $R_\text{min}$ and $R_\text{max}$ are 
determined as convergence parameters, and their typical value is around $n=300-1000$, 
$R_\text{min}=0.1$~bohr and $R_\text{max}=30-100$~bohr.

\clearpage
\subsection{Mass-correction curves and non-adiabatic corrections to the rovibrational energies of 
H$_2^+$ in its ground electronic state, $\tilde{X}\ ^2$\Sgp}
\begin{figure}
  \includegraphics[scale=1.]{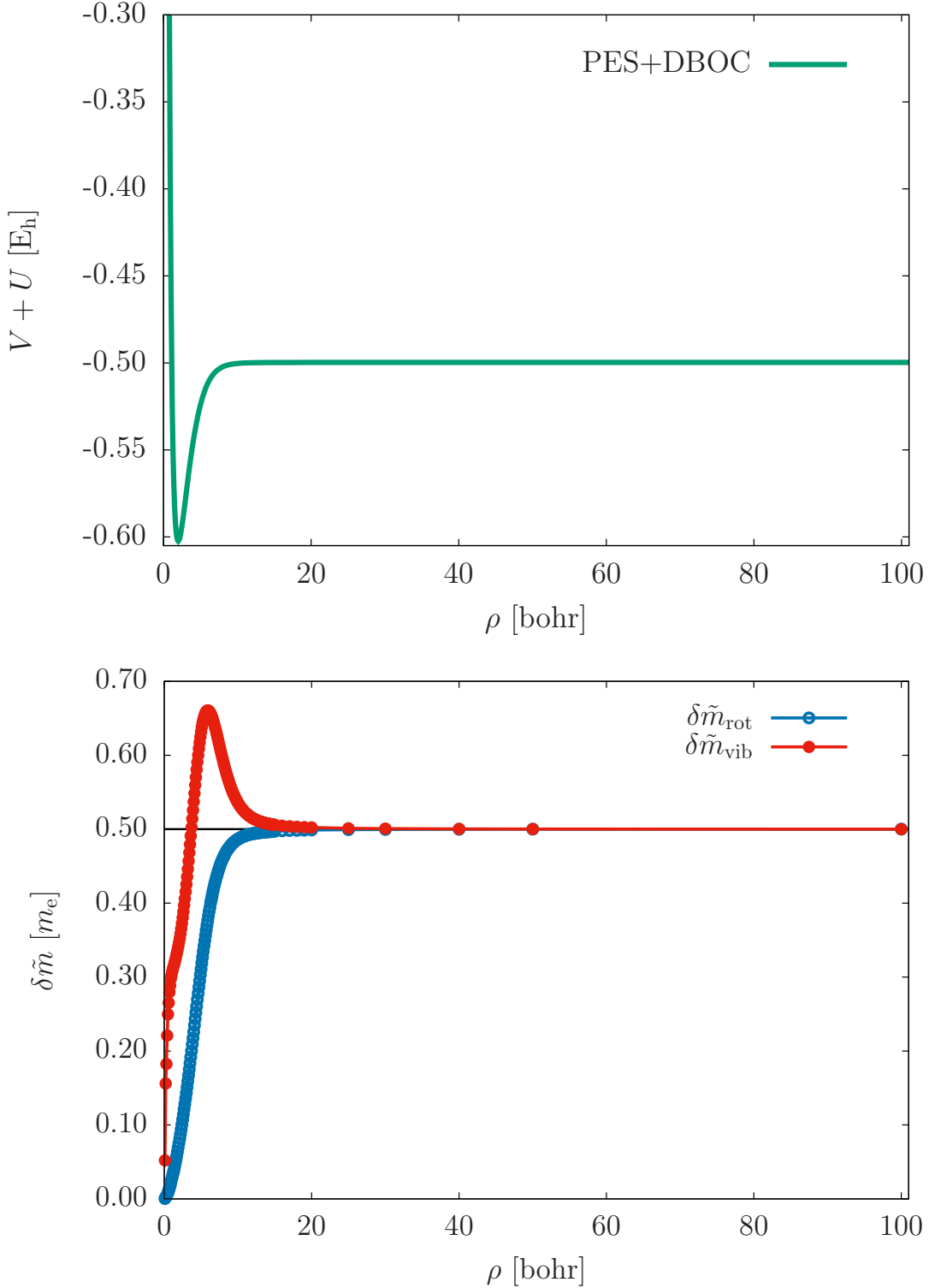} \\
  \caption{%
    H$_2^+$ molecular ion in its
    $\tXdSgp$ electronic state (ground state):
    mass correction functions to the vibrational, 
    $\delta\tilde{m}_\text{vib}=\mmx^\rho\,_\rho$, 
    and the rotational, 
    $\delta\tilde{m}_\text{rot}=\mmx^\theta\,_\theta=\mmx^\phi\,_\phi$ 
    degrees of freedom when the kinetic energy operator is expressed in
    spherical polar coordinates, Eq.~(\ref{eq:tamxcurvdiat}).
    For the definition of the effective rotational and vibrational mass in
    diatomic molecules see 
    Eqs.~(\ref{eq:effvibmassgen})--(\ref{eq:effrotmass}).
    The thin line takes the constant value $1/2=0.5\ m_\text{e}$, 
    which corresponds to an equal share of the single electron between
    the two atomic nuclei.
    \label{fig:masscorrh2px}
  }
\end{figure}

Using the variational method with fECG basis functions and the computational
procedure described in Section~\ref{ch:elstruct}
we have computed the mass-correction functions for the H$_2^+$ molecular ion
in its $\tilde{X}\ ^2$\Sgp\ (ground) electronic state over the 
interval of the $\rho$ internuclear distance.
The resulting rotational and vibrational mass-correction curves computed 
in the present work for H$_2^+$
are visualized in Figure~\ref{fig:masscorrh2px} 
(the computed values are deposited in the \som). 
The figure also shows 
the adiabatic potential energy curve (including DBOC) 
to allow a visual comparison of the improtant features of the mass-correction 
and the potential energy curve.

In Table~\ref{tab:H2pnonaddev} we present the non-adiabatic corrections obtained with 
these mass-correction curves in comparison with Moss's results \cite{Mo93,Mo96}. 
(The non-adiabatic correction to a rovibrational state
is defined as the difference of the eigenvalues obtained with the BO Hamiltonian
with the potential energy curve including the DBOC
minus the corresponding eigenvalues obtained using the full non-adiabatic
Hamiltonian.)
Excellent agreement is observed over all bound and long-lived resonance states. 
(Note that the disagreement for the $(v,J)=(0,2)$ and $(10,4)$
states remains unexplained and might be due to a typographical error 
in the earlier references.)

As to the convergence of the presented results, 
the electronic state, $\psi$, used to compute the mass-correction curves, had 
an energy eigenvalue $E$ converged to better than 0.1~$\mu$\Eh. 
The auxiliary basis set (constructed from functions of $^2$\Sgp\ and $^2$\Pup\ symmetries) 
used to build the $\mx{F}$ matrix, Eq.~(\ref{eq:Fmx}),
was compiled using the parameters optimized for the ground-state energy.
The most stable and accurate mass-correction values were obtained by
using a complex $\varepsilon$ value ($\varepsilon=\text{i}\cdot 10^{-6}$) 
for computing the resolvent in Eq.~(\ref{eq:Fxb}).
This setup was sufficient to converge the non-adiabatic correction energies accurate
to at least $<0.005$~\cm.

In order to assess the necessary accuracy of the electronic energy and wave function 
to obtain accurate non-adiabatic rovibrational energies, 
we have observed that the mass correction functions (and hence, non-adiabatic rovibrational
energies) were robust with respect to inaccuracies in the electronic state: 
a 10--20~$\mu$\Eh\ error of the electronic energy introduced a less than 0.5~\%\
error in the mass-correction values, which had an almost negligible ($<0.005$~\cm) effect on 
the non-adiabatic correction energies.
At the same time, it was found to be important that a sufficiently 
large auxiliary basis set is used in particular for computing the vibrational
mass-correction function at large internuclear separations.

Using the computed, rigorous mass-correction curves, it will be interesting to
compare the (second-order) non-adiabatic rovibrational bound and resonance states
with their three-particle variational counterparts recently computed by Korobov \cite{Ko18},
and the recently measured shape resonances of the hydrogen molecular ion and its deuterated
isotopologue \cite{BeMe16,BeMe16b}. This will offer a unique opportunity to directly test
the numerical accuracy of second-order non-adiabatic theory with respect to 
the numerically exact few-particle variational 
results. Such a direct comparision has recently become available in the literature 
for the rotational states of the hydrogen molecule \cite{PaKo18}.

\clearpage
\begin{table}
  \caption{%
    Non-adiabatic corrections for the rovibrational states
    of H$_2^+$: comparison (deviation) from Moss' results  
    (computed with $m_\text{p}/m_\text{e}=1836.152701$ and $1~\text{E}_\text{h}=219474.63067$~\cm)
    \cite{Mo93,Mo96}.
    \label{tab:H2pnonaddev}
  }
  \begin{rotate}{90}
    \hspace{-22.cm}
    \scalebox{0.72}{%
    \begin{tabular}{@{} c rrrrr rrrrr  rrrrr rrrrr  r @{}}
      \cline{1-22}\\[-0.4cm]
      \cline{1-22}\\[-0.4cm]
$v$/$J$	&	
\multicolumn{1}{c}{0}	&	
\multicolumn{1}{c}{1}	&	
\multicolumn{1}{c}{2}	&	
\multicolumn{1}{c}{3}	&	
\multicolumn{1}{c}{4}	&	
\multicolumn{1}{c}{5}	&	
\multicolumn{1}{c}{6}	&	
\multicolumn{1}{c}{7}	&	
\multicolumn{1}{c}{8}	&	
\multicolumn{1}{c}{9}	&	
\multicolumn{1}{c}{10}	&	
\multicolumn{1}{c}{11}	&	
\multicolumn{1}{c}{12}	&	
\multicolumn{1}{c}{13}	&	
\multicolumn{1}{c}{14}	&	
\multicolumn{1}{c}{15}	&	
\multicolumn{1}{c}{16}	&	
\multicolumn{1}{c}{17}	&	
\multicolumn{1}{c}{18}	&	
\multicolumn{1}{c}{19}	&	
\multicolumn{1}{c}{20}	\\
\cline{1-22}\\[-0.4cm]
0	&	$-0.002$	&	$-0.002$	&	0.059$^\text{a}$	&	$-0.002$	&	$-0.001$	&	$0.000$	&	$-0.001$	&	$-0.001$	&	$-0.002$	&	$-0.002$	&	$-0.004$	&	$-0.002$	&	$-0.003$	&	$-0.003$	&	$-0.002$	&	$-0.002$	&	$-0.003$	&	$-0.003$	&	$-0.002$	&	$-0.002$	&	$-0.003$	\\
1	&	$0.000$	&	$-0.002$	&	$-0.003$	&	$-0.002$	&	$-0.002$	&	$0.000$	&	$-0.001$	&	$-0.002$	&	$-0.002$	&	$-0.002$	&	$-0.002$	&	$-0.003$	&	$-0.001$	&	$-0.002$	&	$-0.002$	&	$-0.002$	&	$-0.003$	&	$-0.002$	&	$-0.003$	&	$-0.003$	&	$-0.003$	\\
2	&	$-0.002$	&	$-0.002$	&	$-0.003$	&	$-0.001$	&	$-0.002$	&	$-0.001$	&	$-0.003$	&	$-0.002$	&	$-0.001$	&	$-0.002$	&	$-0.002$	&	$-0.002$	&	$-0.002$	&	$-0.003$	&	$-0.003$	&	$-0.003$	&	$-0.003$	&	$-0.002$	&	$-0.002$	&	$-0.003$	&	$-0.003$	\\
3	&	$-0.002$	&	$-0.003$	&	$-0.002$	&	$-0.002$	&	$-0.002$	&	$-0.002$	&	$-0.003$	&	$-0.001$	&	$-0.002$	&	$-0.003$	&	$-0.003$	&	$-0.002$	&	$-0.002$	&	$-0.002$	&	$-0.003$	&	$-0.003$	&	$-0.003$	&	$-0.003$	&	$-0.003$	&	$-0.003$	&	$-0.003$	\\
4	&	$-0.002$	&	$-0.002$	&	$-0.002$	&	$-0.003$	&	$-0.002$	&	$-0.003$	&	$-0.002$	&	$-0.002$	&	$-0.002$	&	$-0.002$	&	$-0.003$	&	$-0.003$	&	$-0.003$	&	$-0.003$	&	$-0.003$	&	$-0.002$	&	$-0.003$	&	$-0.003$	&	$-0.003$	&	$-0.003$	&	$-0.002$	\\
5	&	$-0.002$	&	$-0.003$	&	$-0.003$	&	$-0.003$	&	$-0.003$	&	$-0.002$	&	$-0.002$	&	$-0.003$	&	$-0.002$	&	$-0.003$	&	$-0.003$	&	$-0.002$	&	$-0.003$	&	$-0.003$	&	$-0.003$	&	$-0.003$	&	$-0.002$	&	$-0.003$	&	$-0.003$	&	$-0.003$	&	$-0.002$	\\
6	&	$-0.003$	&	$-0.002$	&	$-0.002$	&	$-0.003$	&	$-0.002$	&	$-0.003$	&	$-0.003$	&	$-0.003$	&	$-0.002$	&	$-0.003$	&	$-0.003$	&	$-0.003$	&	$-0.002$	&	$-0.002$	&	$-0.003$	&	$-0.003$	&	$-0.003$	&	$-0.003$	&	$-0.003$	&	$-0.003$	&	$-0.003$	\\
7	&	$-0.003$	&	$-0.003$	&	$-0.004$	&	$-0.003$	&	$-0.003$	&	$-0.002$	&	$-0.003$	&	$-0.004$	&	$-0.003$	&	$-0.003$	&	$-0.002$	&	$-0.003$	&	$-0.003$	&	$-0.003$	&	$-0.003$	&	$-0.003$	&	$-0.002$	&	$-0.003$	&	$-0.003$	&	$-0.002$	&	$-0.002$	\\
8	&	$-0.002$	&	$-0.003$	&	$-0.003$	&	$-0.003$	&	$-0.003$	&	$-0.003$	&	$-0.003$	&	$-0.003$	&	$-0.003$	&	$-0.003$	&	$-0.003$	&	$-0.003$	&	$-0.003$	&	$-0.003$	&	$-0.002$	&	$-0.003$	&	$-0.002$	&	$-0.002$	&	$-0.002$	&	$-0.002$	&	$-0.002$	\\
9	&	$-0.003$	&	$-0.003$	&	$-0.003$	&	$-0.003$	&	$-0.003$	&	$-0.002$	&	$-0.002$	&	$-0.002$	&	$-0.003$	&	$-0.003$	&	$-0.003$	&	$-0.002$	&	$-0.002$	&	$-0.002$	&	$-0.002$	&	$-0.003$	&	$-0.002$	&	$-0.002$	&	$-0.001$	&	$-0.001$	&	$-0.002$	\\
10	&	$-0.003$	&	$-0.002$	&	$-0.003$	&	$-0.003$	&	$-0.273$$^\text{b}$	&	$-0.003$	&	$-0.002$	&	$-0.003$	&	$-0.003$	&	$-0.002$	&	$-0.002$	&	$-0.002$	&	$-0.002$	&	$-0.002$	&	$-0.002$	&	$-0.002$	&	$-0.002$	&	$-0.001$	&	$-0.001$	&	$-0.001$	&	$-0.002$	\\
11	&	$-0.002$	&	$-0.002$	&	$-0.003$	&	$-0.003$	&	$-0.002$	&	$-0.002$	&	$-0.002$	&	$-0.003$	&	$-0.002$	&	$-0.003$	&	$-0.002$	&	$-0.001$	&	$-0.002$	&	$-0.001$	&	$-0.001$	&	$-0.001$	&	$-0.002$	&	$-0.002$	&	$-0.001$	&	$-0.001$	&	$-0.001$	\\
12	&	$-0.002$	&	$-0.002$	&	$-0.002$	&	$-0.002$	&	$-0.002$	&	$-0.003$	&	$-0.002$	&	$-0.002$	&	$-0.003$	&	$-0.002$	&	$-0.001$	&	$-0.001$	&	$-0.002$	&	$-0.002$	&	$-0.002$	&	$-0.001$	&	$-0.001$	&	$-0.001$	&	$-0.001$	&	\multicolumn{1}{c}{$\bullet$}	&		\\
13	&	$-0.002$	&	$-0.002$	&	$-0.002$	&	$-0.002$	&	$-0.002$	&	$-0.002$	&	$-0.002$	&	$-0.002$	&	$-0.002$	&	$-0.002$	&	$-0.002$	&	$-0.002$	&	$-0.001$	&	$-0.001$	&	$-0.001$	&	$-0.001$	&	$-0.001$	&	\multicolumn{1}{c}{$\bullet$}	&		&		&		\\
14	&	$-0.001$	&	$-0.002$	&	$-0.002$	&	$-0.002$	&	$-0.001$	&	$-0.001$	&	$-0.002$	&	$-0.001$	&	$-0.001$	&	$-0.002$	&	$-0.001$	&	$-0.001$	&	$-0.001$	&	$0.000$	&	$-0.001$	&	\multicolumn{1}{c}{$\bullet$}	&		&		&		&		&		\\
15	&	$-0.001$	&	$-0.002$	&	$-0.001$	&	$-0.002$	&	$-0.001$	&	$-0.001$	&	$-0.001$	&	$-0.001$	&	$-0.001$	&	$-0.001$	&	$-0.001$	&	$-0.001$	&	$0.002$	&	\multicolumn{1}{c}{$\bullet$}	&		&		&		&		&		&		&		\\
16	&	$-0.001$	&	$-0.001$	&	$-0.001$	&	$-0.002$	&	$-0.001$	&	$-0.001$	&	$-0.001$	&	$-0.001$	&	$-0.001$	&	$0.001$	&	\multicolumn{1}{c}{$\bullet$}	&		&		&		&		&		&		&		&		&		&		\\
17	&	$-0.001$	&	$0.000$	&	$-0.001$	&	$-0.001$	&	$-0.001$	&	$-0.001$	&	$0.000$	&	\multicolumn{1}{c}{$\bullet$}	&		&		&		&		&		&		&		&		&		&		&		&		&		\\
18	&	$0.000$	&	$0.000$	&	$0.000$	&	$0.000$	&	\multicolumn{1}{c}{$\bullet$}	&		&		&		&		&		&		&		&		&		&		&		&		&		&		&		&		\\
19	&	$0.000$	&	$0.000$	&		&		&		&		&		&		&		&		&		&		&		&		&		&		&		&		&		&		&		\\
%
\cline{1-22}\\[-0.4cm]
$v$/$J$	&	
\multicolumn{1}{c}{21}	&	
\multicolumn{1}{c}{22}	&	
\multicolumn{1}{c}{23}	&	
\multicolumn{1}{c}{24}	&	
\multicolumn{1}{c}{25}	&	
\multicolumn{1}{c}{26}	&	
\multicolumn{1}{c}{27}	&	
\multicolumn{1}{c}{28}	&	
\multicolumn{1}{c}{29}	&	
\multicolumn{1}{c}{30}	&	
\multicolumn{1}{c}{31}	&	
\multicolumn{1}{c}{32}	&	
\multicolumn{1}{c}{33}	&	
\multicolumn{1}{c}{34}	&	
\multicolumn{1}{c}{35}	&	
\multicolumn{1}{c}{36}	&	
\multicolumn{1}{c}{37}	&	
\multicolumn{1}{c}{38}	&	
\multicolumn{1}{c}{39}	&	
\multicolumn{1}{c}{40}	&	
\multicolumn{1}{c}{41}	\\
\cline{1-22}\\[-0.4cm]
0	&	$-0.003$	&	$-0.003$	&	$-0.003$	&	$-0.002$	&	$-0.003$	&	$-0.003$	&	$-0.004$	&	$-0.003$	&	$-0.003$	&	$-0.004$	&	$-0.003$	&	$-0.004$	&	$-0.004$	&	$-0.004$	&	$-0.004$	&	$-0.003$	&	$-0.003$	&	$-0.003$	&	$-0.003$	&	$-0.001$	&	\multicolumn{1}{c}{$\bullet$}	\\
1	&	$-0.003$	&	$-0.003$	&	$-0.003$	&	$-0.003$	&	$-0.003$	&	$-0.003$	&	$-0.003$	&	$-0.003$	&	$-0.003$	&	$-0.003$	&	$-0.003$	&	$-0.003$	&	$-0.004$	&	$-0.003$	&	$-0.003$	&	$-0.003$	&	$-0.003$	&	$0.006$	&	\multicolumn{1}{c}{$\bullet$}	&		&		\\
2	&	$-0.002$	&	$-0.004$	&	$-0.003$	&	$-0.003$	&	$-0.004$	&	$-0.003$	&	$-0.003$	&	$-0.003$	&	$-0.003$	&	$-0.004$	&	$-0.003$	&	$-0.003$	&	$-0.003$	&	$-0.003$	&	$-0.003$	&	$-0.002$	&	$-0.193$$^\text{e}$	&	\multicolumn{1}{c}{$\bullet$}	&		&		&		\\
3	&	$-0.003$	&	$-0.002$	&	$-0.003$	&	$-0.003$	&	$-0.003$	&	$-0.003$	&	$-0.003$	&	$-0.003$	&	$-0.003$	&	$-0.003$	&	$-0.002$	&	$-0.002$	&	$-0.002$	&	$-0.002$	&	$0.047$$^\text{d}$	&	\multicolumn{1}{c}{$\bullet$}	&		&		&		&		&		\\
4	&	$-0.003$	&	$-0.004$	&	$-0.003$	&	$-0.003$	&	$-0.003$	&	$-0.003$	&	$-0.003$	&	$-0.002$	&	$-0.002$	&	$-0.002$	&	$-0.002$	&	$-0.002$	&	$0.003$	&	\multicolumn{1}{c}{$\bullet$}	&		&		&		&		&		&		&		\\
5	&	$-0.003$	&	$-0.002$	&	$-0.003$	&	$-0.002$	&	$-0.002$	&	$-0.002$	&	$-0.002$	&	$-0.002$	&	$-0.002$	&	$-0.002$	&	$0.006$	&	\multicolumn{1}{c}{$\bullet$}	&		&		&		&		&		&		&		&		&		\\
6	&	$-0.002$	&	$-0.003$	&	$-0.002$	&	$-0.002$	&	$-0.002$	&	$-0.002$	&	$-0.002$	&	$-0.001$	&	$-0.002$	&	\multicolumn{1}{c}{$\bullet$}	&		&		&		&		&		&		&		&		&		&		&		\\
7	&	$-0.002$	&	$-0.002$	&	$-0.002$	&	$-0.001$	&	$-0.002$	&	$-0.002$	&	$-0.002$	&	$0.018$$^\text{c}$	&	\multicolumn{1}{c}{$\bullet$}	&		&		&		&		&		&		&		&		&		&		&		&		\\
8	&	$-0.002$	&	$-0.002$	&	$-0.001$	&	$-0.001$	&	$-0.001$	&	$0.005$	&	\multicolumn{1}{c}{$\bullet$}	&		&		&		&		&		&		&		&		&		&		&		&		&		&		\\
9	&	$-0.001$	&	$-0.001$	&	$-0.001$	&	$-0.001$	&	\multicolumn{1}{c}{$\bullet$}	&		&		&		&		&		&		&		&		&		&		&		&		&		&		&		&		\\
10	&	$-0.001$	&	$-0.001$	&	\multicolumn{1}{c}{$\bullet$}	&		&		&		&		&		&		&		&		&		&		&		&		&		&		&		&		&		&		\\
11	&	\multicolumn{1}{c}{$\bullet$}	&		&		&		&		&		&		&		&		&		&		&		&		&		&		&		&		&		&		&		&		\\
\cline{1-22}\\[-0.4cm]
\cline{1-22}\\
\multicolumn{22}{l}{$^\text{a,b}$ The stability and estimated accuracy of our correction terms is much better than the order of magnitude of the deviation.} \\
\multicolumn{22}{l}{$^\text{c,d,e}$ The accuracy of these results has been labelled in the present work as well as Ref.~\cite{Mo93}. The symbol $\bullet$ is used for states which have been predicted } \\
\multicolumn{22}{l}{by Moss without giving their energy and non-adiabatic correction explicitly in Refs.~\cite{Mo93,Mo96}.}\\      
    \end{tabular}
    }
  \end{rotate}
\end{table}

\clearpage
\section{Summary and conclusions}
General curvilinear expressions have been derived 
for the second-order non-adiabatic kinetic energy operator.
The derivations have been carried out using the general Jacobian and metric tensors
of the coordinate transformation, thereby they are in direct connection
with the numerical-kinetic energy operator approach (with constant masses)
also used in the GENIUSH protocol \cite{MaCzCs09}.
While for the constant-mass case one has to transform the ``div grad'' operator 
to curvilinear coordinates, 
we had to consider the ``div $\massmx$ grad'' operator with the $\massmx$ 
\emph{coordinate-dependent} \emph{tensor} quantity within the operator.
As a result, should a general mass-correction tensor surfaces become available for polyatomic
molecules, their implementation 
in the (polyatomic) rovibrational GENIUSH program has been made straightforward.

At the moment, we are not aware of any widely available electronic structure package which
we could use to compute the mass-correction tensor over a wide range of nuclear configurations
of polyatomic molecules
(note however that Ref.~\cite{prx17} points into a very promising direction).
Hence, we have modified (simplified) our in-house preBO code 
originally developed for the computation of 
isolated few-particle quantum systems \cite{MaHuMuRe11a,MaHuMuRe11b,MaRe12,Ma13,Ma18}
to few-electron systems being in interaction with external charges (nuclei) and 
use floating explicitly-correlated
Gaussian functions for the spatial part of the basis set. 
We report the first numerical results using this computational setup
and present the rigorous mass-correction values for
the H$_2^+$ molecular ion in its ground electronic state over the $[0.1,100]$~bohr
range of the internuclear distance. 
Using these mass-correction functions in the 
rovibrational computations we reproduce Moss' non-adiabatic
corrections \cite{Mo93,Mo96} for H$_2^+$ within a few 0.001~\cm\ 
for the bound and long-lived resonance states. 
In a forthcoming paper \cite{Ma18He2p} 
further applications of the developed formalism and methodology 
are presented for the $^4$He$_2^+$ molecular ion in its ground electronic state 
for which a highly accurate potential energy and DBOC curve is available \cite{TuPaAd12}. 
It is observed that full account of the rigorous mass correction functions substantially reduce the 
deviation of experimental \cite{CaPyKn95,SeJaMe16} and computed rotational(-vibrational) energy intervals.

\vspace{1cm}
\noindent\textbf{Supplementary Material}\\ 
The Supplementary Material contains 
non-adiabatic mass correction values for H$_2^+$ ($\tilde X\ ^2\Sigma_\text{g}^+$).

\vspace{1cm}
\noindent\textbf{Acknowledgment}\\ 
Financial support of the Swiss National Science Foundation through
a PROMYS Grant (no. IZ11Z0\_166525) is gratefully acknowledged. 
The author is thankful to Krzysztof Pachucki for discussions about non-adiabatic perturbation theory, 
and to Federica Agostini for the explanation of
the mass-correction tensor derivation starting from exact factorization.

\clearpage

\end{document}